 \documentclass [twocolumn,trackchanges]{aastex63}
\usepackage[caption=false]{subfig}
\usepackage{savesym}
\savesymbol{tablenum}
\usepackage{siunitx}
\restoresymbol{SIX}{tablenum}
\usepackage{fontawesome}
\usepackage{multirow}
\usepackage{graphicx}
\usepackage{booktabs}
\usepackage{comment}
\usepackage{etoolbox}
\usepackage{amsmath}
\newcommand{\ie}{i.\,e. }

\newcommand{\eg}{e.\,g. }

\newcommand{\xedit}[1]{{#1}}
\received{June 1, 2019}
\revised{January 10, 2019}
\accepted{\today}
\submitjournal{ApJ}

\shorttitle{Inpainting foreground maps}
\shortauthors{Puglisi et al.}
\graphicspath{{./}{figures/}}
\begin{document}

\title{ Inpainting Galactic Foreground Intensity and Polarization maps using \\ 
Convolutional Neural Networks}

\author{Giuseppe Puglisi }
\email{gpuglisi@lbl.gov}
\affiliation{Department of Physics, Stanford University, Stanford, CA 94305, USA}
\affiliation{Kavli Institute for Particle Astrophysics and Cosmology, SLAC National Accelerator Laboratory, Menlo Park, CA 94025, USA}
\affiliation{Computational Cosmology Center, Lawrence Berkeley National Laboratory, Berkeley, CA 94720, USA}

\author{ Xiran Bai }
\affiliation{Department of Physics, Stanford University, Stanford, CA 94305, USA}
\affiliation{Kavli Institute for Particle Astrophysics and Cosmology, SLAC National Accelerator Laboratory, Menlo Park, CA 94025, USA}
%\affiliation{Department of Physics, Yale University, New Haven, CT 06520, USA}



\begin{abstract}
%250 word limit 
\xedit{The Deep Convolutional Neural Networks (DCNNs) have been a popular tool for image generation and restoration. 
In this work, we applied DCNNs to the problem of inpainting non-Gaussian astrophysical signal}, in the context of Galactic diffuse emissions at the millimetric and sub-millimetric regimes, specifically  Synchrotron and Thermal Dust emissions. Both signals are affected by contamination at small angular scales due to extra-galactic radio sources (the former)  and dusty star-forming galaxies (the latter). 
\xedit{We compare the performance of the standard diffusive inpainting with that of two novel methodologies relying on DCNNs, namely Generative Adversarial Networks and Deep-Prior. We show that the methods based on the DCNNs are able to reproduce the statistical properties of the ground truth signal more consistently with a higher confidence level.} The Python Inpainter for Cosmological and AStrophysical SOurces (\textsc{PICASSO})   is a package encoding a suite of  inpainting methods described in this work and  has been made publicly available. 

\end{abstract}


\section{Introduction} \label{sec:intro}

\noindent Over the  last few years, the use of machine learning techniques has become increasingly popular in analyzing scientific data. In particular, the use of the Deep Convolutional Neural Networks (DCNNs) has opened a wide range of interesting applications \citep{farsian2020foreground,2019A&C....2800307C,Aylor2019,Krachmalnicoff2019,perraudin2019cosmological,2018ComAC...5....4R,Mustafa17}.

\xedit{In this work, we investigate how the problem of estimating and reconstructing missing or masked regions of observations can be better solved using DCNNs.} These techniques have been widely used for image restoration and face completion and in principle can be similarly applied to generate semantic content for  astrophysical signals.

In particular, we focus on the case of reconstructing polarized signal emitted in the radio and sub-mm regimes:  i)  at $\nu \lesssim \SI{60}{ \GHz}$ where the emission is mostly dominated by Galactic synchrotron, described by  a power law $\beta_{synch}\sim -3 $ \citep{2018A&A...618A.166K},  ii) at $ \nu \gtrsim \SI{150}{\GHz}$  where most of the polarization is due to the thermal Galactic dust grains aligning with the Galactic magnetic field, described by a \emph{modified blackbody}  law \citep{planck2018}.  {Emission coming from molecules (like Carbon Monoxide) and from anomalous microwave emission are expected to be polarized at this regime of frequencies, although their degree of polarization is expected to be  lower (few per cent, see for details \citet{Puglisi2016a,Dickinson2018})  compared  to that of  dust and synchrotron (about $10 \%$).} 

At $80 <\nu< 110 \, \si{\giga\hertz} $, the Cosmic Microwave Background (CMB) polarization has a non-negligible contribution especially at high Galactic latitudes.  A reliable assessment of  both synchrotron and dust polarized  emissions in the two regimes i) and ii)  is critical to separate  the Galactic contamination in CMB measurements and further detect the divergence-less  pattern in the CMB polarization called $B$-mode. CMB B-modes, at degree angular scales, are directly related to the imprint  of  a stochastic background of gravitational waves produced during the inflationary phase of our universe, commonly referred as \emph{tensorial}  anisotropies.
To date,  \emph{primordial} $B$-modes have not yet been detected  and the  latest  upper  limits  have been provided by \citet{2019arXiv191005748S, 2019arXiv191002608A, 2018PhRvL.121v1301B}. Future experiments aim at better characterizing diffuse polarized emission from our own Galaxy with high-sensitivity measurements \citep{2019BAAS...51g.209C, 2019JCAP...02..056A}.

At the arcminute angular scales, B-modes are sourced by  the gravitational lensing of large scale structures which    deflect  the CMB \emph{scalar} polarization anisotropies into  the so-called \emph{lensing} B-modes, (see  latest constraints  in \citet{2019arXiv191005748S, 2017ApJ...848..121P, 2017JCAP...06..031L}).   At these scales,  extra-galactic radio sources and star-forming galaxies are the major polarized contaminants. The majority of these contaminants mostly appears as bright and unresolved  point-sources in a typical CMB map ( latest measurements can be found in \citet{gupta2019,2019MNRAS.486.5239D}). 
 \citet{puglisi2018} have shown that  hundreds of polarized sources will be detected by the forthcoming experiments given the expected nominal sensitivity and the observation sky fraction  ( $\sim 10- 30 \% $). Hence an aggressive masking may be applied on maps surveyed by the forthcoming CMB experiments, preventing  a high-resolution  Galactic foreground template as well as a reliable analysis involving high-order estimators  beyond the two-point  correlation function.  Reconstructing signals in the masking area to fill the missing data is done to ameliorate these issues, a procedure sometimes referred to as \emph{inpainting} (used in e.g. \citet{Starck_2013}). 
 
 In this work, three different methodologies are tested to \emph{inpaint}  maps at the locations of extra-galactic point-sources. 
 Two of the inpainting techniques involve \emph{generative} DCNNs. We compare the DCNNs inpainting performances with the standard diffusive inpainting approach used in \citet{2016JCAP...05..055B}, which is simply filling the missing pixel with the average value of its nearest-neighbours. 

We organize the paper as follows:  In Sect.\ref{sec:methods}, we present the three inpainting methodologies adopted in this work. Sect.\ref{sec:data} describes the data used for training and validation purposes. Finally, Sect.\ref{sec:results} includes the results achieved by inpainting on simulations (Subsect.\ref{subsec:fidelity}) and on more realistic data-sets (Subsect.\ref{subsec:nulls}). Finally, we apply our inpainting method to the the  map of  the S-band Polarization All Sky Survey (SPASS, \citet{2019MNRAS.489.2330C}) at several source locations (Sect.\ref{subsec:spass})  and demonstrate that we robustly recover the background signal.  

\section{Methods of inpainting} \label{sec:methods}

\noindent Inpainting algorithms can be divided into two main groups: i) diffusive-based methods and ii) learning-based methods that rely on training DCNNs to fill the missing pixels with the predictions  learned from a training data-set. 
We choose three inpainting techniques from both groups: Sect.\ref{subsec:nn} describes  a diffusive-based method from group i), and  Sect. \ref{subsec:dp} and \ref{subsec:ca} present  methods from group ii). 

\subsection{Nearest-Neighbors} \label{subsec:nn}  
 \noindent One of the simplest  inpainting methods is the\emph{ diffusive inpainting} described in \citet{2016JCAP...05..055B}, which has been adopted in \citet{2014, 2016}.  In this method, each masked pixel is iteratively filled with the mean value of  its nearest-neighbor pixels, being  often referred to as the Nearest-Neighbors (NN) in the image reconstruction algorithm.
 
 The iterative procedure can be performed in two ways: i) \emph{Gauss-Seidel} method, which computes the average of neighbors at the current iteration. {As a consequence, the pixels near the boundary are updated in earlier iterations while pixels near the center of the inpainting regions require several iterations.} ii) \emph{Jacobi} method,  which estimates the average value from a buffer of pixel values at the previous iteration. 
 \citet{2016JCAP...05..055B}    found that  $\sim \mathcal{O}(10^3) $   iterations were   needed  to inpaint   $\sim 10\  \, \si{arcmin}$ areas on a map with $\sim 2\ \, \si{arcmin}$ pixel size.
 Although \citet{2016JCAP...05..055B} found that both methods did not impact the quality of the inpainted results, the Gauss-Seidel method achieves faster convergence than the Jacobi method. We therefore adopted the former method as suggested by \citet{2016JCAP...05..055B}.  
 
 % include  a set of pictures with inpainted patches . 
 
\subsection{Deep-Prior}\label{subsec:dp}
\noindent    { Deep-Prior (DP;\ \citealt{deeprior}) is the first methodology encoding DCNNs we use in this work. The fundamental  assumption of DP is that the information required to reconstruct an image are essentially encoded in the  input image itself (which might be already corrupted, noisy or with missing pixels) and in the network architecture used for the reconstruction. 
It has the peculiarity of  being an \emph{untrained} network and the  inpainting procedure can be summarized as:  i) process the image to be inpainted  through the  convolutional layers   for several iterations (usually more than 1,000),  ii) fit  for the neural network  parameters (or \emph{weights}), and iii) evaluate  a  \emph{loss} function for a given set of parameters.  \citet{deeprior} proposed several loss functions specifically related to the task to be performed (e.g. super-resolution, inpainting, de-noising).  }

 { 
Inpainting can be  formalized as an \emph{image generating} {procedure}, $f_{\theta} $,   mapping from the  so called latent space $\mathbb{R} ^{m} $ into the feature space $\mathbb{R}^n$, with $n=N_{pix} \times N_{pix}$ being the size of the input images and with  parameters $\theta$. Generative networks takes as an input a random vector $z$, known as the \emph{the prior} and  outputs an image  $ \tilde{x}  = f_{\theta}(z)$ given $z$ and the  parametrization. Generally,  the whole network architecture is  symmetric and  can be structured into two main parts: an \emph{encoder}  which represents the input into a lower dimensional space  and a \emph{decoder}  meant to do the opposite, upsample  from the low dimensional space to the original input one.  }

 {We therefore build the DP architecture by  following the prescription given in  \citet{deeprior} for the task of  inpainting\footnote{Further details can be found in supplementary material, \url{https://dmitryulyanov.github.io/deep_image_prior}. }, and  building  a \emph{U-Net} type architecture sketched in Fig.\ref{fig:dp_arch} and  summarized in Table \ref{tab:dp_arch} and \ref{tab:dp2}.  } 

   { The optimal set of parameters   $\theta^*$  is obtained  minimizing the  loss function given a ground-truth   image $x_0$,  with missing pixels correspondent to a binary mask $m $ (with 0 in the masked region and 1 elsewhere)  : 
 \begin{equation}
     E(x^*, x_0) = \parallel \left( f_{\theta^*}(z) -x_0 \right) \odot m\parallel^2 , \label{eq:lossDP}
 \end{equation}
 with  $x^* = f_{\theta^*}(z)$ being the output of the inpainting reconstruction procedure and $\odot$  the Hadamard's product. The minimization process of    eq. (\ref{eq:lossDP}) is performed with  the \emph{gradient descent} algorithm. Notice that this loss  {function} does not depend on the values of the missing pixels, but only on the features outside the mask where the ground-truth is known.}

\subsubsection{Deep-Prior architecture }
\begin{figure*}[!htpb]
    \centering
    \includegraphics[width=2\columnwidth]{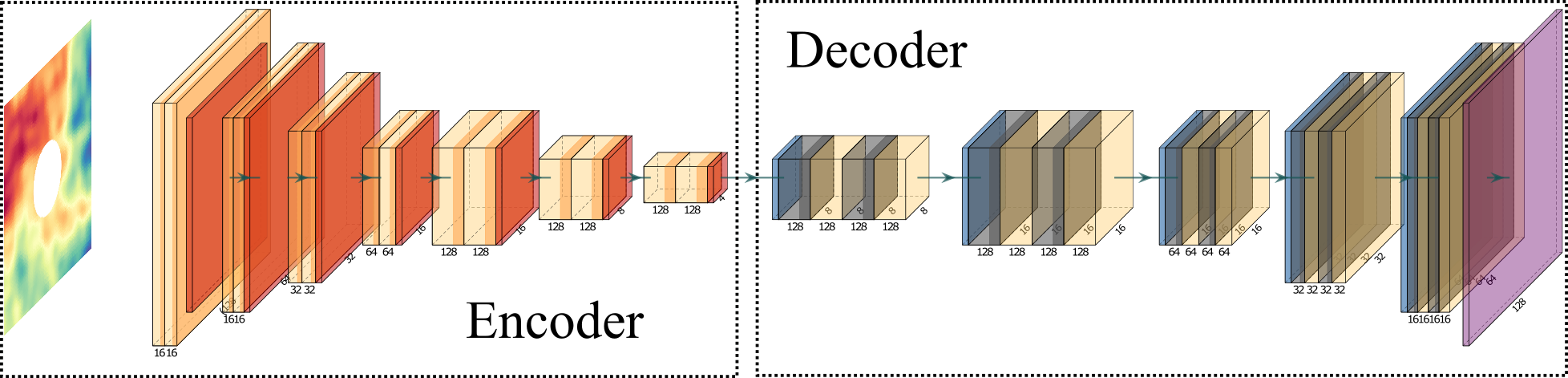}
    \caption{Sketch of the DP architecture. \xedit{This U-Net type architecture consists of a series of convolutional blocks: the encoder and the decoder blocks. The inputs first pass the encoder blocks for down-sampling, and then are up-sampled through the decoder blocks for the final outputs. Further details of the architecture and parameters can be found in Table \ref{tab:dp_arch}.}}
    \label{fig:dp_arch}
\end{figure*}

  { The architecture of DP is sketched  in Fig.\ref{fig:dp_arch} , the network read the  inputs  are $128\times128$ images of Galactic foreground maps and they are processed through a series of 6 convolutional blocks aimed at down-sampling the images  (with a kernel size of $k_d=3$) and an increasing number of channels, $n_d$ (ranging from 16 to 128 ). Each block includes two convolutional layers respectively  with stride $s=1$  and $s=2$  followed by the application of the  \emph{Leaky Residual Exponential Linear Unit (LeakyRELU) } activation, with a learning rate parameter of $\alpha_{LR} =0.1 $. At the end of each down-sampling block there is  another convolutional layer with stride $s=1$  followed by  another LeakyRELU. The decoder blocks  follow the exact symmetry of the encoder ones and differ only in the kernel size adopted $k_u = 5$ . The structure of each  up-sampling block, include  a  convolution layer  and  LeakyRELU repeated twice and teminates with an  upsampling layer increasing the  size of the output by factor of  $2$.  In total, the DP network has 4,237,441 parameters. }  {
The whole architecture is implemented using \textit{Keras}\footnote{ \url{https://keras.io/api/}} package with TensorFlow backend \footnote{ \url{https://www.tensorflow.org/}}.  } 

\subsection{Generative Adversarial Networks with Contextual Attention}\label{subsec:ca}

\noindent   {Generative Adversarial Networks (GAN; \citealt{Goodfellow2014}) is a popular machine learning algorithm useful in several contexts, especially in image reconstruction. The overall GAN framework can be synthesized as an  \emph{adversarial interplay} between two networks, a \emph{generator}, $G$, and a \emph{discriminator}, $D$. \xedit{$G$ is trained to generate fake samples that resemble the training set, and $D$ is responsible for judging whether a given sample is generated by $G$ or is a real image that belongs to the training set.} \xedit{The loss of the network is designed such that $G$ will get penalized for producing images that looks fake (statistically deviate from the training set), and $D$ will get penalized for misjudging the originality of the images. The two networks are then trained until the generator is able to generate good quality images that the discriminator cannot tell if they are generated or not. }}

  {In the context of image reconstruction, GAN seeks for coherency between generated and existing pixels by adopting a convolutional encoder-decoder generator network (similar to  the one described in Sect.\ref{subsec:dp}). As a result, one of the biggest advantages of using GAN based methods is that they are less affected by problems observed in simple convolutional networks such as boundary artifacts and blurry textures, making the reconstructed images inconsistent with the surrounding regions \citet{IizukaSIGGRAPH2017, yu2015multiscal}. } 

 {
Recently, \citet{generative} presented a novel generative inpainting procedure based on GAN with an extra-branch in the architecture aiming at providing more coherence in the reconstructed area given features in the uncorrupted regions of the image,  referred here as the \emph{contextual attention} branch. The proposed generator network can be then structured into two stages: a \emph{coarse reconstruction}  stage which  employs a generator built with an encoder-decoder architecture  trained to inpaint the missing region, and a \emph{refinement} stage aiming at improving the generator with local and global   features estimated with contextual attention. }  \xedit{The first stage, the coarse stage, is based on an encoder-decoder network to roughly generate the content in the missing region. The second stage, the refinement stage, is organized into two parallel convolutional pathways both fed with the output of the coarse stage},  { \ie the full image with an approximated content in the missing region. One of the two branches aims at hallucinating novel contents in the missing region in order to better refine the content inside the mask by injecting smaller  scale  features. The other branch encodes the contextual attention to enhance  spatial coherency of the local features inside the masked area with the global features.} To better visualize how the contextual attention works, we report in Fig.\ref{fig:colormap} an example of inpainting with GAN and the attention map related to this case.  The attention map helps to identify  which regions of the input image  the  contextual-attention  focuses on in order to refine the corrupted image.   
 
 { The output of the refinement stage is  then combined and fed to a single decoder for the final inpainting output. } Both the coarse and refinement stages of the generator network are trained \emph{end-to-end} with reconstruction losses. We follow the prescriptions in \citet{generative} and  adopt
a weighted sum of pixel-wise $\ell_1$ loss  to train explicitly the coarse  and the refinement networks.  { For the discriminator, we adopted two Wasserstein GAN (WGAN) adversarial losses \citep{2017arXiv170107875A}  aimed at evaluating separately the global and local features of the generated images.}

\begin{figure*}
    \centering
    \includegraphics[width=2\columnwidth]{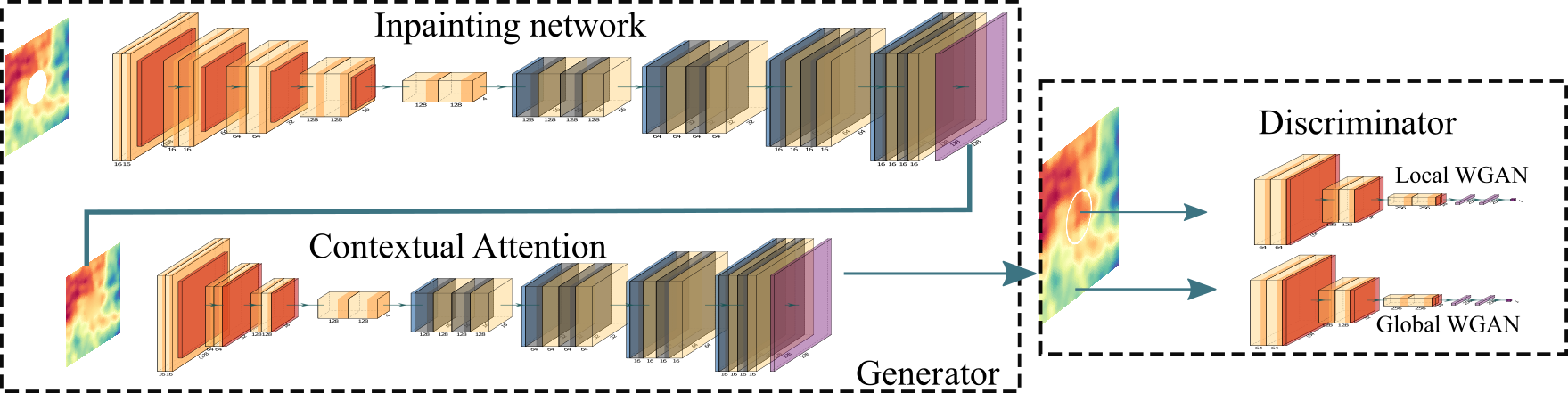}
    \caption{Sketch of the GAN architecture with contextual attention implemented in this work. The Generator network consists of a coarse and a refinement stage which respectively produce a rough and refined reconstruction of the corrupted image. The input of the Discriminator coincides with the output of the refinement stage. Both Generator and Discriminator are trained end-to-end together. Further details of the network architecture and parameters can be found in the Table \ref{table:archi}.}
    \label{fig:gan_arch}
\end{figure*}

\begin{table*}[htpb!]
    \centering
    
    \begin{tabular}{|l l|   }
    \hline  
    \textbf{Deep-Prior } &\textbf{Network} \\
    \hline 
    $n_d$ & \texttt{[16, 32, 64, 128, 128, 128]}\\
    $k_d$& \texttt{[3, 3, 3, 3, 3, 3]} \\
    $n_u$ &  \texttt{[128, 128, 128, 64, 32, 16 ]}  \\
    $k_u$& \texttt{[5, 5, 5, 5, 5, 5]} \\
    \hline 
    
    \end{tabular}
    \caption{Architecture for DP  network as described in \citet{deeprior}.  $n_u, n_d$ correspond to respectively the number of upsampling and downsampling filters, $ k_d,k_u$ correspond to the  kernel sizes. }
    \label{tab:dp_arch}
    
    \begin{tabular}{|l l c|l l c | }
    \hline  
     \textbf {Encoder Block} &$s$  & Parameters& \textbf {Decoder Block} &$s$  & Parameters \\
      \hline
    \texttt{Conv2D}      & 1 & $k_d[i] \times k_d[i] \times n_d[i] \times 1 $& \texttt{Conv2D}   &  1 &$k_u[i] \times k_u[i] \times n_u[i] \times 1 $\\
    \texttt{Conv2D}        & 2 & $k_d[i] \times k_d[i] \times n_d[i] \times 1 $& \texttt{LeakyRELU} &&$\alpha_{LR}=0.1 $ \\
    \texttt{LeakyRELU} &&$\alpha_{LR}=0.1 $ & \texttt{Conv2D}   &  1 &$k_u[i] \times k_u[i] \times n_u[i] \times 1 $ \\
    \texttt{Conv2D}        & 1 & $k_d[i] \times k_d[i] \times n_d[i] \times 1 $& \texttt{LeakyRELU} &&$\alpha_{LR}=0.1 $ \\
    \texttt{LeakyRELU} &&$\alpha_{LR}=0.1 $ &\texttt{UpSampling2D}   & &$\mathtt{size }= (2, 2) $ \\
    \hline 
    \end{tabular}
    \caption{Sequence of layers encoded in the $i$-th  encoder and  decoder  block adopted for DP. $s$ refers to the stride size.}\label{tab:dp2}
    
\end{table*}

\begin{table*}[htpb!]
    \centering
    
    \begin{tabular}{|l l| l|  }
    \hline  
    \textbf{Inpainting Network } &  &\textbf{Contextual-Attention branch }\\
    \hline 
    $n_d $& \texttt{[32,64,64,128,128,128,128}&\texttt{L1:[32,64,64,128,128,128,128]}\\
    &\texttt{128,128,128,128,128]} &\texttt{L2:[128,128]}\\
        $k_d$& \texttt{[5,3,3,3,3,3,3,3,3,3,3]} &\texttt{L1:[5,3,3,3,3,3], L2:[3,3]} \\
    $s_d$ & \texttt{[1,2,1,2,1,1,1,1,1,1,1]}&\texttt{L1:[1,2,1,2,1,1], L2:[1,1]} \\
    $D$ & \texttt{[1,1,1,1,1,2,4,8,16,1,1]}& \\
    $n_u$ &  \texttt{[128,64,64,32,16]} &\texttt{[128,64,64,32,16]} \\
    $k_u$&\texttt{[5,5,5,5,5]}& \texttt{[5,5,5,5,5]} \\
    $s_u$ & \texttt{[1,1,1,1,1]}&\texttt{[1,1,1,1,1]}\\
    
    \hline 
    \textbf{Local WGAN } &  &\textbf{Global WGAN }\\
    \hline 
    $n$ & \texttt{[64,128,256,512]}&\texttt{[64,128,256,256]}\\ 
    $k$ &\texttt{[5,5,5,5]}&\texttt{[5,5,5,5]} \\ 
    $s$ &\texttt{[2,2,2,2]}&\texttt{[2,2,2,2]} \\ 
    \hline 
    \end{tabular}
    \caption{Architecture and parameters used for GAN as described in \citet{generative}.  For each network, we indicate with  $n, k,s$  respectively the number of channels, the kernel size and  the stride size. In particular, for the inpainting and contextual attention networks we label with $d$ ($u$) the sizes for the  downsampling (upsampling) of   filters. $D$ refers to the dilation rate. \texttt{L1} and \texttt{L2} in the Contextual-Attention column refer to the two parallel encoders in the refinement stage.  }
    \label{table:archi}
    \end{table*}

\subsubsection{GAN  architecture }
\begin{figure}
    \centering
    \includegraphics[width=1\columnwidth]{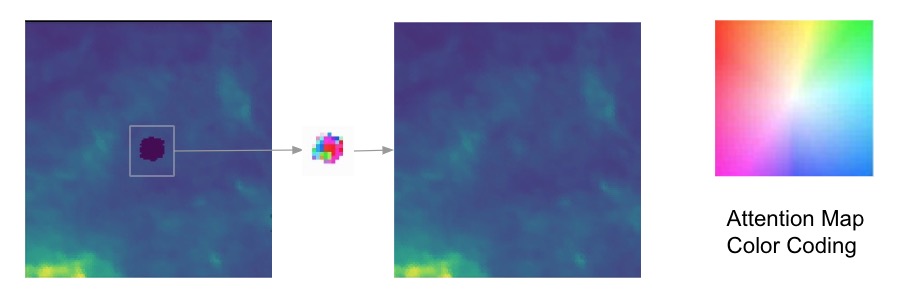}
    \caption{Visualization of an inpainting contextual-attention map.  {The colors in the color coding indicate the portion of the whole image the neural network has focused on in order to inpaint a given pixel location in the masked area.}  In this particular example, most of the pixels are inpainted by GAN by looking at the upper and lower left region of the image}
    \label{fig:colormap}
\end{figure}

 { 
 The overall architecture used in generative inpainting is shown  in Fig.\ref{fig:gan_arch} ( details for convolutional layers can be found in Table \ref{table:archi}). Moreover, we choose to adopt the same hyper-parameters in setting the network as    the one provided in \cite{generative}.  Each convolutional layer is implemented by using   mirror padding, without batch normalization and with Exponential Linear Units (ELUs, \citet{clevert2015fast}) activation functions.  We summarize into 4 sub-networks the GAN  implementation in this work: 
 \begin{itemize}
     \item \textbf{Inpainting network}   performs   the coarse stage of the Generator network, and presents an encoder-decoder architecture made of 16  blocks of convolutional layers. For the downsampling part, we choose $k=3$ for all the kernels, except for the 1st one ($k=5$), and set the stride  to $s=2$ only   for  the  2nd  and 4th convolutional block (for the rest $s=1 $). Starting from the 6th  to the 9th block \emph{dilation}  is also applied to the convolution  with an increasing dilation rate from 2 to 16. The application of dilations  aims at   capturing  more global features from the input without increasing the size of parameters. 
     The upsampling part encodes two upsampling layers after the 12th and the 14th block, for all the convolutions we choose $k_u=5$ and $s_u=1$. The number of channels are reported in Table \ref{table:archi}.
    \item \textbf{Contextual attention branch} is aimed at refining the  coarsely  inpainted image output from the previous network. This branch has a very similar architecture as the inpainting one with the main difference that  it is made of    two parallel encoders concatenated to a single decoder.  One of the encoders estimates the contextual attention. In this branch, the attention map (Fig.\ref{fig:colormap}) is estimated. The architecture details are listed in Table\ref{table:archi} and  we refer to \texttt{L1} and \texttt{L2} for the two   encoders. 
\item \textbf{Local and Global critics} in the Discriminator network aim at labeling whether the  inpainted  image is coherent inside and outside the masked area. Both critics are made of 4 blocks with $k=5$ and $s=2$ and with an increasing number of channels $n$ from $64$ to $512$ ($256$) for the local (global) network. The last convolution is linked onto a fully connected layer and to the WGAN loss function. 
\end{itemize}
}

\section{Data and Simulations} \label{sec:data}

In this study, we mainly focus on inpainting maps of two  emissions in the  microwave regimes: \ie  synchrotron and thermal dust.   These two  foreground components contaminate the CMB polarization measurements and need to be modeled both at large and small angular scales for foreground removal. Since the statistical properties of Galactic foregrounds are highly non-Gaussian, an interesting  application of DCNNs is to reconstruct images with complex, non-Gaussian features. In contrast, traditional inpainting methods used in CMB studies such as the Gaussian Constrained Inpainting methods \citep{hoffman91, Bucher_2012} assume that the background is a Gaussian field, making it incapable of capturing the non-Gaussianity in the foregrounds.

Both Galactic  {foregrounds}, unpolarized  and polarized emissions data  (respectively encoded in the brightness temperature, T  and \emph{Stokes } parameters Q and U maps) are simulated using the PySM package \citep{2017MNRAS.469.2821T} and will be described below in Subsections \ref{subsec:sync} and \ref{subsec:dust}.

\subsection{Galactic Synchrotron}\label{subsec:sync}
For the synchrotron data, we consider   SPASS \citet{2019MNRAS.489.2330C} which observed the Southern sky ($\delta <-1^{\circ} $) at 2.3 GHz with an 8.9 arcmin full width at half maximum (FWHM).   The methodology used to generate the intensity and polarization maps are described in \citealt{2019MNRAS.489.2330C}.
98.6\% of the pixels in the Q and U SPASS maps {have} signal-to-noise ratio (SNR) $>3$, making these maps a  promising synchrotron polarization template \citep{2018A&A...618A.166K}. \citet{2016ApJ...829....5L} cross-matched the extra-galactic radio quasars (mostly steep spectrum sources) detected by SPASS with the ones detected by  NRAO/VLA Sky Survey, (NVSS, \citet{1998AJ....115.1693C}), at $1.4 $ GHz and released a polarization catalogue {with} 533 bright sources in the overlapping area of the two surveys. However, because the  SPASS T, Q and U maps are filled with radio sources, an assessment on the map level to the smallest angular scales   is essentially  compromised by the point-source bias. Therefore, masking and inpainting these sources provide an overall benefit in fully exploiting the angular scales probed by SPASS. 

In order to inpaint  {the} SPASS map, we firstly create a simulated training  data-set   which will be used for training the GAN (training set) and for  evaluating the quality of the reconstructions (testing set).  We therefore  simulate SPASS synchrotron-only  TQU maps  at the SPASS frequency with  \texttt{s1} PySM  model. This model is  one of the most representative  since it parametrizes the synchrotron power-law  emission with  a spatially varying spectral index. Maps are pixellized {on a } HEALPix \footnote{ \url {https://healpix.sourceforge.io} }\citep{2005ApJ...622..759G, 2019JOSS....4.1298Z} {\texttt{nside=2048} grid}  convolved with a $8.9 $ arcmin FWHM beam. 

\subsection{ Thermal Dust }\label{subsec:dust}
 We use  the thermal dust maps at 353 and 857 GHz from the third \emph{Planck} public release\footnote{\url{https://pla.esac.esa.int}}.   Both frequency maps are dominated by the thermal dust emission emitted by our own Galaxy and  encode   contribution from Cosmic Infrared Background (CIB). We choose the 353 GHz frequency channel  in order to test the  inpainting techniques on both dust temperature and polarization maps. Because the  857 GHz channel is not polarization sensitive, we use this  channel  to assess how the different SNR and  different CIB contribution affect  inpainting reconstructions in total intensity. 
 {At these frequencies, the emission is dominated by star forming galaxies, and blazars are expected to have minor contribution to the total intensity.}
 
 We build the training set   by simulating TQU Thermal dust maps at 353 GHz with the \texttt{d1} PySM model, which describes the modified blackbody emission law with a spectral index and a temperature, both spatially varying. Maps are simulated  on a \texttt{nside=2048} grid and convolved with a 5 arcmin FWHM  beam, similarly to the one  in the \emph{Planck} 353 GHz observations. 
 
 Furthermore, since {the pixel values of the maps} are rescaled during the training and  inpainting processes, the GAN reconstruction is not affected by the overall amplitude of a signal. We will show in  Sec.\ref{subsec:nulls} that inpainting performances do not change as a function of frequency as long as the brightest signal in the map coincides with the one used in the training.   This   { independence of frequency  } is the reason why we trained the GAN for thermal dust emission { using PySM signal-only simulations with single frequency} at 353 GHz.

\subsection{The  Training  Data-set}
Both the training and testing data-sets are made from the PySM simulated maps. 
We forecast with PS4C package \footnote{\url {https://gitlab.com/giuse.puglisi/PS4C}  } \citep{puglisi2018},  the number of sources, $N_{\rm src}$ whose density fluxes will be  detected at  $5\sigma$  significance above the  sensitivity flux. For a generic large aperture forthcoming CMB  experiment  (\eg  \citet {SO2019}), the forecasted detections is about $N_{\rm src} \sim \num{30000} $. We  then generate a point-source mask by randomly extracting  $N_{src}$ locations following a \emph{Poisson} distribution. We mask the sources with circular holes centered at the source locations and with a radius three times larger than the  beam FWHM size (namely $26.7 $ and $15$ arcmin respectively  for synchrotron and dust maps). 

Both  masked  and  unmasked  maps  are then  split into  $3\times 3 \, \si{\deg^2}$ square tiles, composed of $128  \times 128$ pixels  with  resolution closer to the HEALPix  one (\ie $\sim 1.5$  arcmin at \texttt{nside=2048}). We finally build the training set for the GAN network by combining   $\num{45000}$  images from   square patches extracted equally from T, Q and U maps. 
The remaining $  \num{ 5000}$ images are used for validation and $500$ for  testing. 

\section{Results} \label{sec:results}

\begin{figure*}[htpb!]

\subfloat[  ]{
\begin{minipage}{2\columnwidth}
 \includegraphics[width=1\columnwidth, trim=0cm 3.4cm 1.5cm 3.0cm , clip=true]{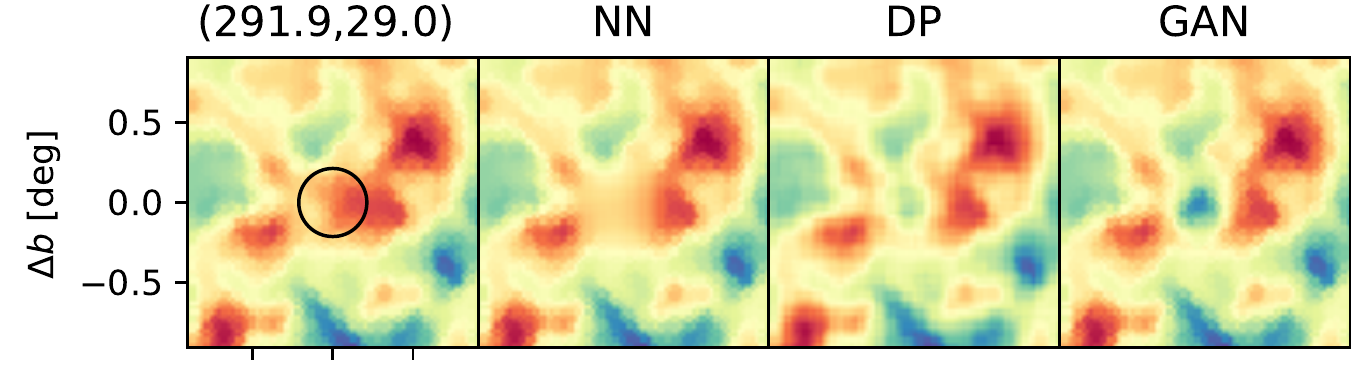}
% \qquad 
 \includegraphics[width=1\columnwidth, trim=0cm 3.4cm 1.5cm 3.0cm , clip=true]{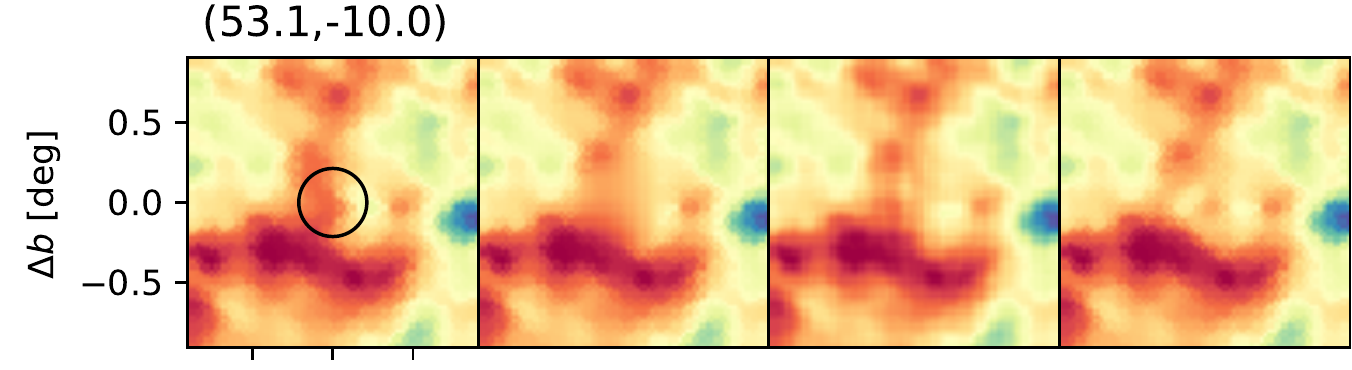}
\end{minipage}
}

\subfloat[  ]{
\begin{minipage}{2\columnwidth}
  \includegraphics[width=1\columnwidth, trim=0cm 3.4cm 1.5cm 3.0cm , clip=true]{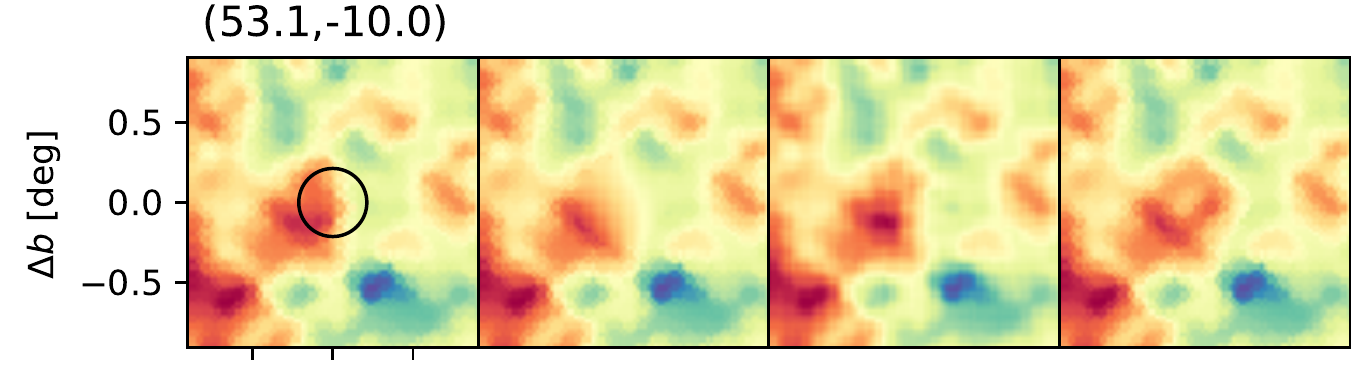}

  \includegraphics[width=1\columnwidth, trim=0cm 2.5cm 1.5cm 3.0cm , clip=true]{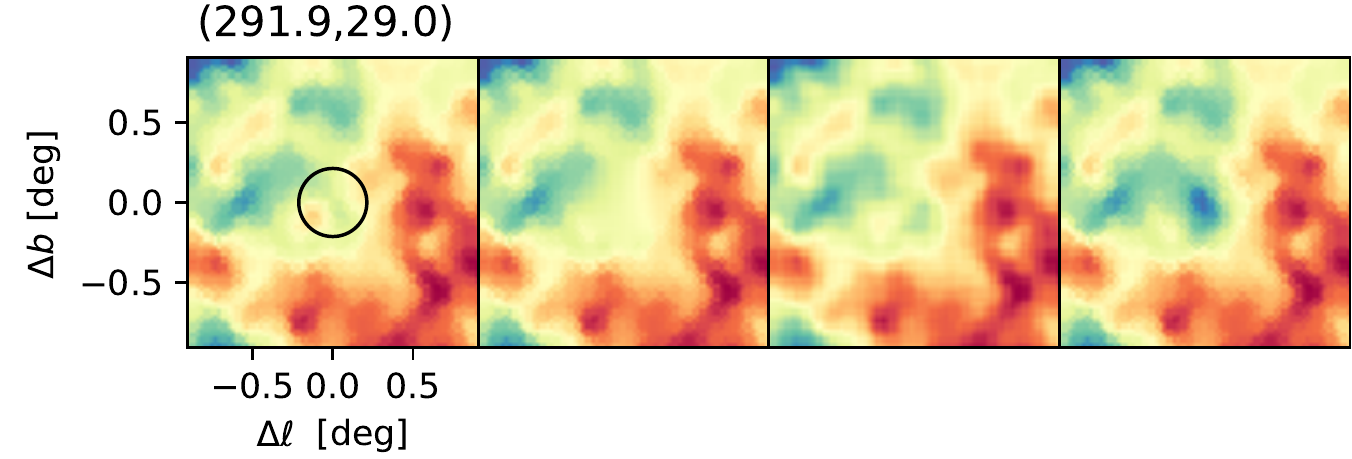}
\end{minipage} 
}
    \caption{Thumbnail  $1.5\times 1.5\,  \si{\deg}^2$ crops for (a)  Q  and (c) U   maps  predictions of thermal dust. The radius of the  reconstructed area ( black circle) is 15 \si{arcmin}. Columns from left to right show  ground truth maps from the test set, predictions obtained with DP, NN and GAN respectively. The colorbar is set to be the same in each row  by \texttt{MinMax} rescaling  all the images with the same \emph{min} and \emph{max} values of the ground-truth ones. Notice that we further zoom in the dust maps to better inspect of the inpainted region (maps are originally $3 \times 3 \, \si{deg}^2$. Temperature maps are shown in  Fig.\ref{fig:inpaint_temp} of the Appendix. }\label{fig:inpaint_dust}
\end{figure*}

\begin{figure*}[htpb!]

 \subfloat[  ]{
\begin{minipage}{2\columnwidth}
\includegraphics[width=1.\columnwidth, trim=.5cm 3.4cm 1.5cm 3.0cm , clip=true]{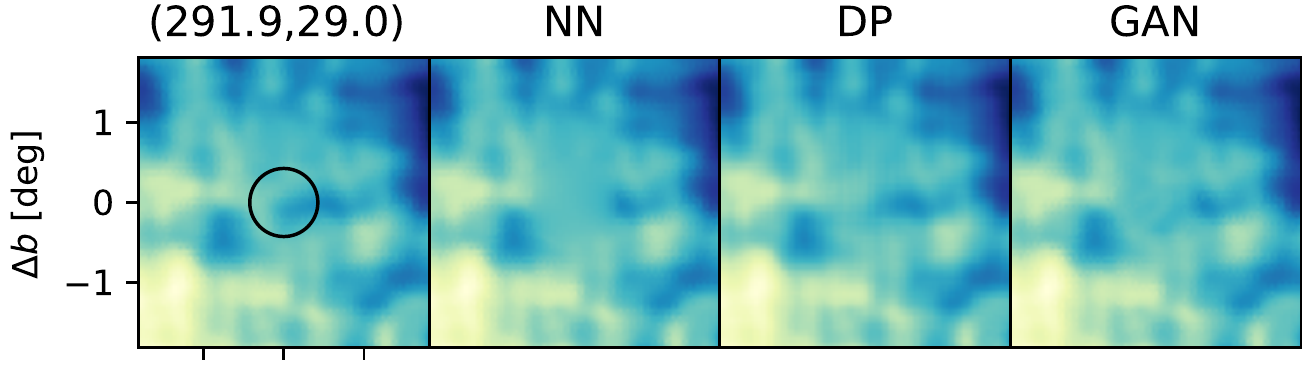}
 \includegraphics[width=1.\columnwidth, trim=.5cm 3.4cm 1.5cm 3.0cm , clip=true ]{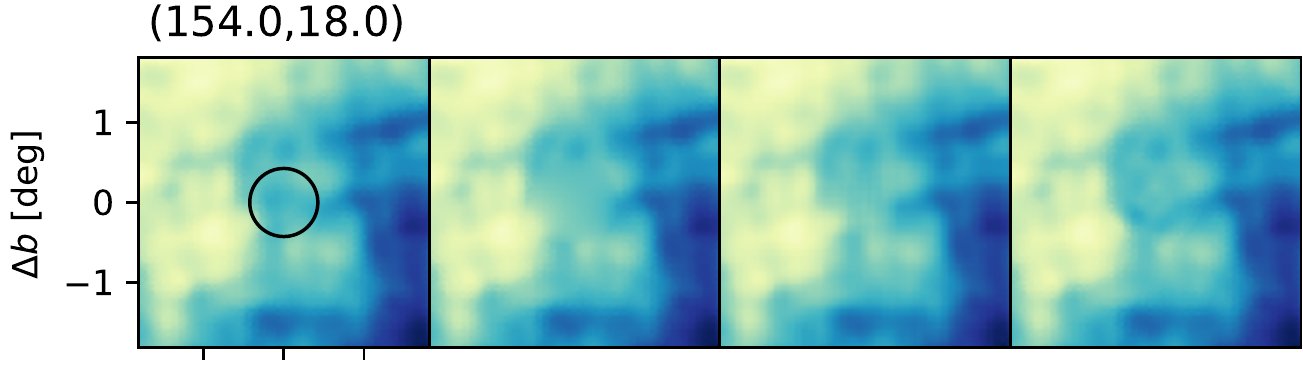}
 \end{minipage}
 }

\subfloat[  ]{
\begin{minipage}{2\columnwidth}
\includegraphics[width=1.\columnwidth, trim=.5cm 3.4cm 1.5cm 3.0cm , clip=true]{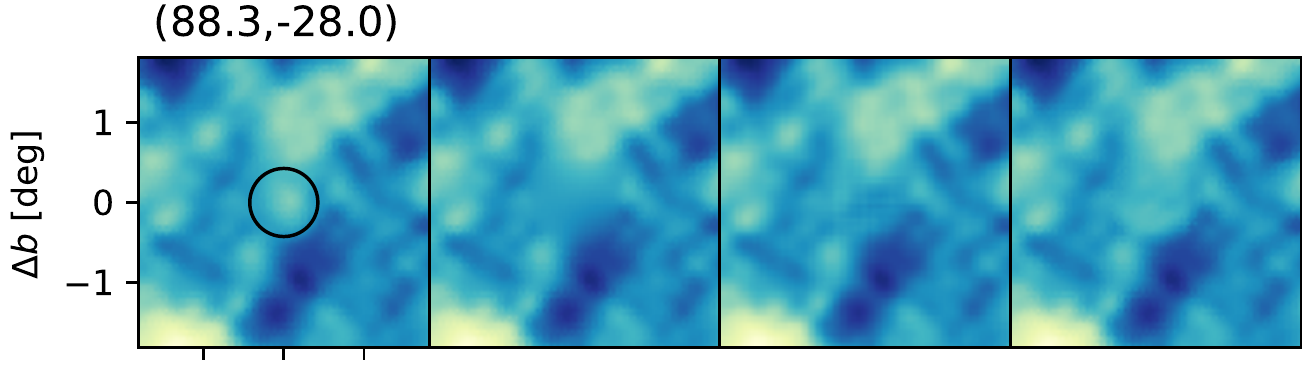}
 \includegraphics[width=1.\columnwidth, trim=.5cm 2.5cm 1.5cm 3.0cm , clip=true ]{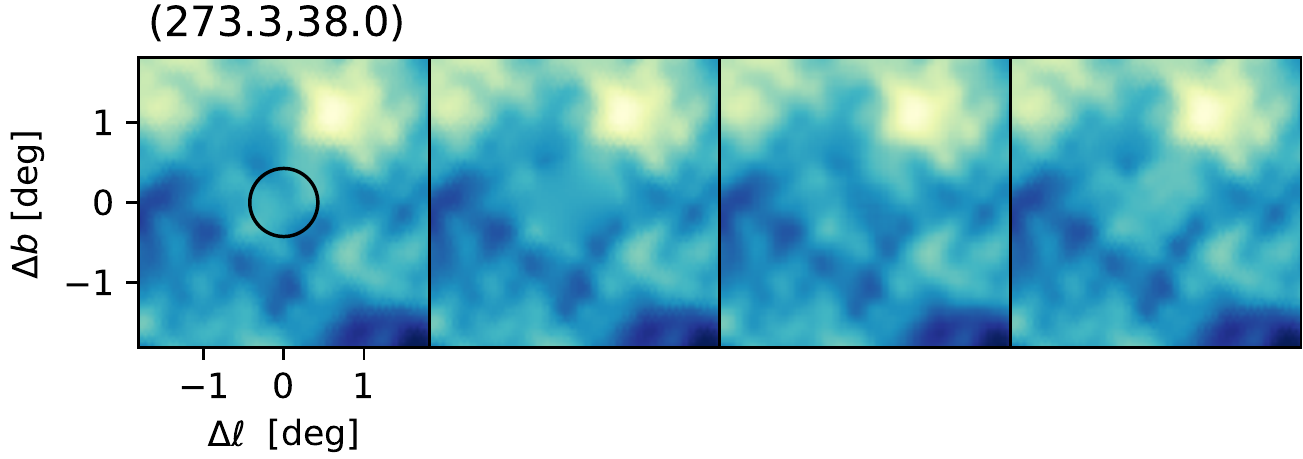}
\end{minipage} 
}
    \caption{Thumbnail  $3\times 3\,  \si{\deg}^2$ crops for (a) Q  and (c) U   maps  predictions of synchrotron emission. The arrangement and the  colorbar setting are the same as those in Fig.\ref{fig:inpaint_dust}. The radius of the  reconstructed area (black circle) is 30 \si{arcmin}. Temperature maps are shown in  Fig.\ref{fig:inpaint_temp}.}
    \label{fig:inpaint_synch}
\end{figure*}

\begin{figure*}[ htpb!]
 \subfloat[  ]{
\begin{minipage}{2\columnwidth}
\includegraphics[width=1\columnwidth, trim=0cm 3.4cm 1.5cm 3.0cm , clip=true]{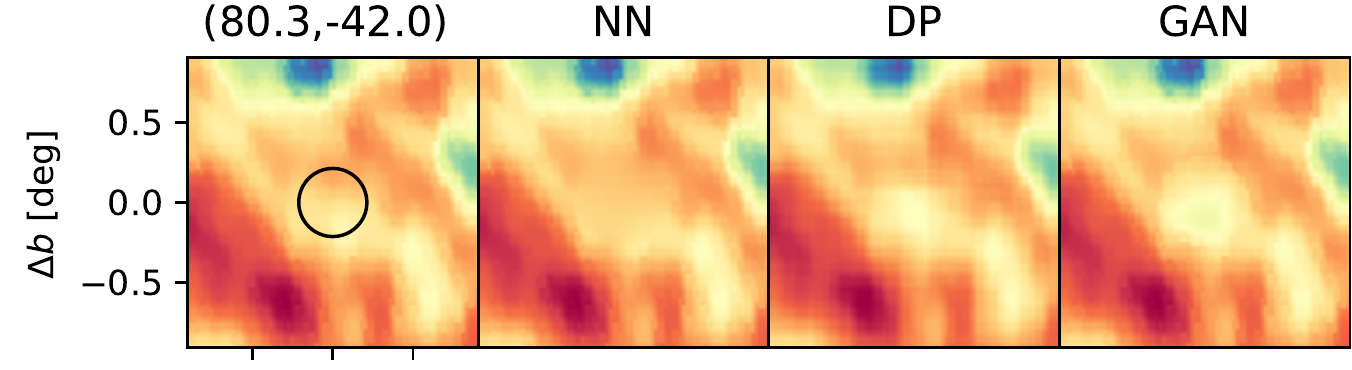} 

\includegraphics[width=1\columnwidth, trim=0cm 3cm 1.5cm 3.0cm , clip=true]{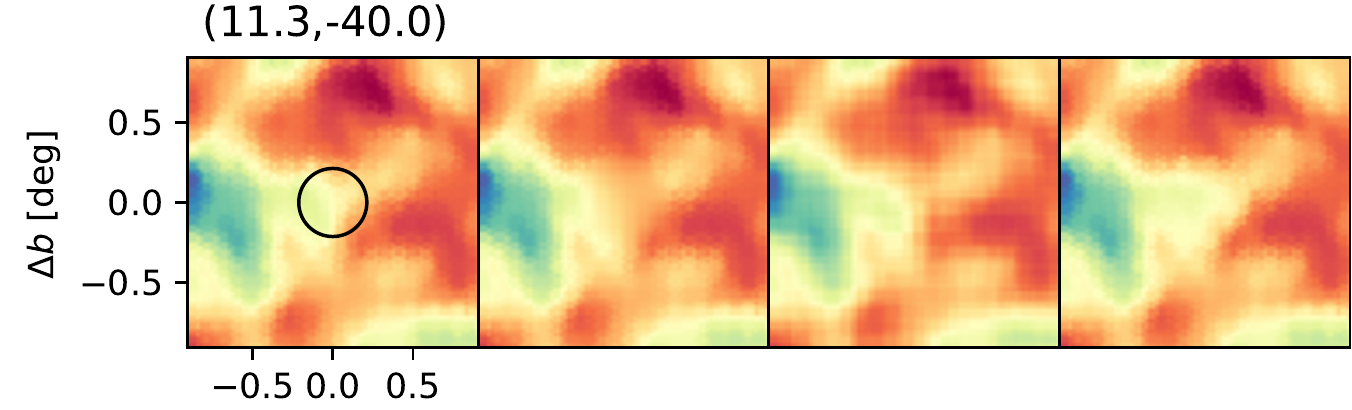} 
\end{minipage}
}

\subfloat[  ]{
\begin{minipage}{2\columnwidth}
    \includegraphics[width=1.0\columnwidth, trim=0cm 3.4cm 1.5cm 3.0cm , clip=true ]{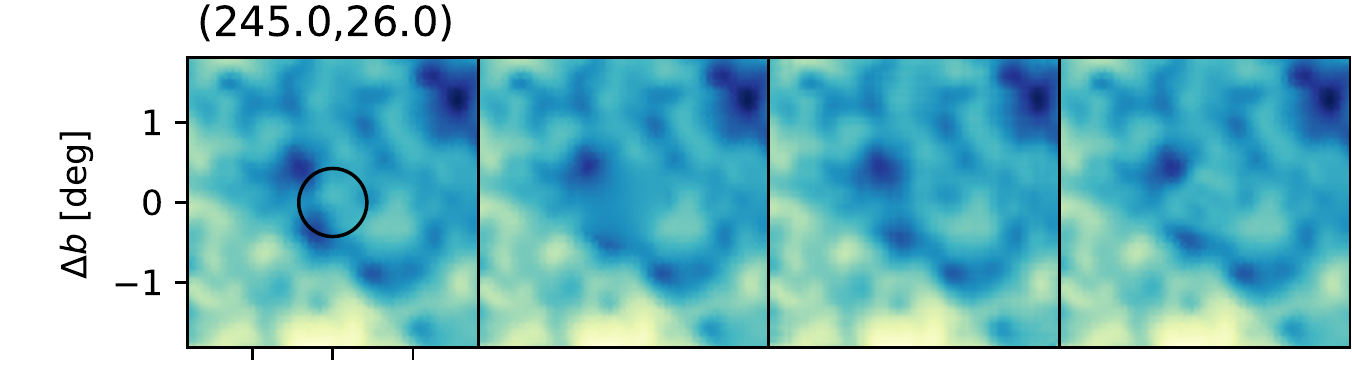}
    
  \includegraphics[width=1.\columnwidth, trim=0cm 2.5cm 1.5cm 3.0cm , clip=true ]{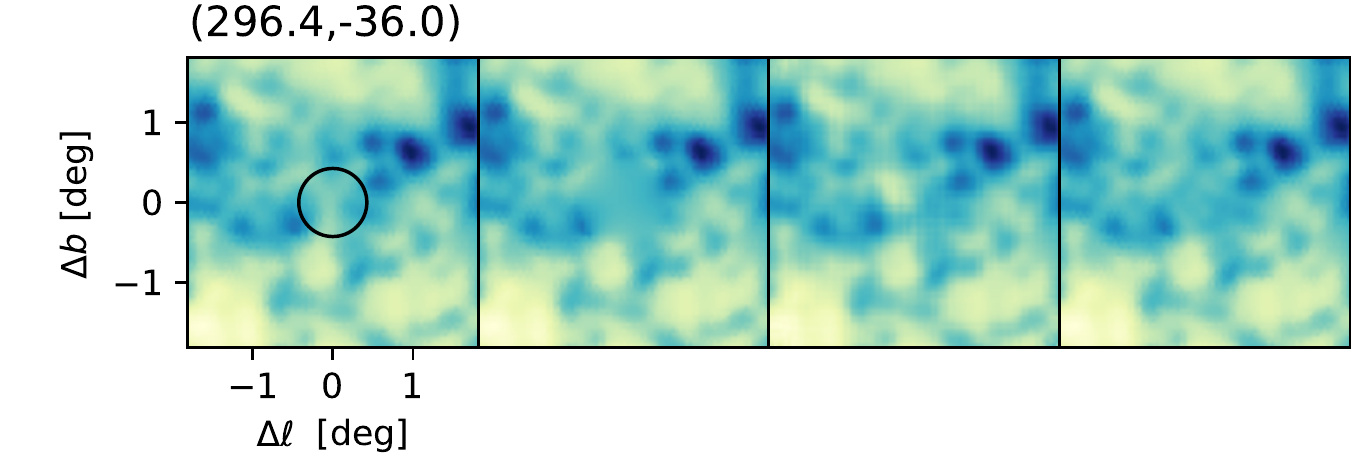}
  \end{minipage}
  }
\caption{  Thumbnail  images for (a)dust, (b)   synchrotron  T  maps  predictions from the testing set simulated with PySM. The radius of the  reconstructed area is shown as grey circle.}
    \label{fig:inpaint_temp}
\end{figure*}

\noindent  Figures \ref{fig:inpaint_dust},  \ref{fig:inpaint_synch} and \ref{fig:inpaint_temp}  show examples of   maps extracted from the test set and  reconstructed with the three methods outlined in Section \ref{sec:methods}. We estimate the minimum and maximum values of each ground-truth image to rescale it (together with the respective inpainted ones )  to $[ 0,1]$ with the  \texttt{MinMax}  normalization. This rescaling forces the generated maps to have the same range as the test ones, so the  differences between the ground-truth and reconstructed map can be spotted more easily. Notice that we further zoom in dust maps ($1.5\times 1.5\,  \si{\deg}^2$ crops) to better inspect the inpainted region.

As expected,  the  inpainting performed with NN algorithm is  smooth and lacks  of finer details,  making  them distinguishable from   the original map.  

On the contrary, for the inpainting performed with DP and GAN,  it is  harder to point out which one is the ground-truth and which is the reconstructed maps. Moreover, we note that DP and GAN are able to reproduce the large scale features, the most correlated  angular scales, as well as the typical  Q/U pattern for the polarization maps.  

  \begin{figure*}[htpb!]
    \centering
   
     {\includegraphics[width=2\columnwidth]{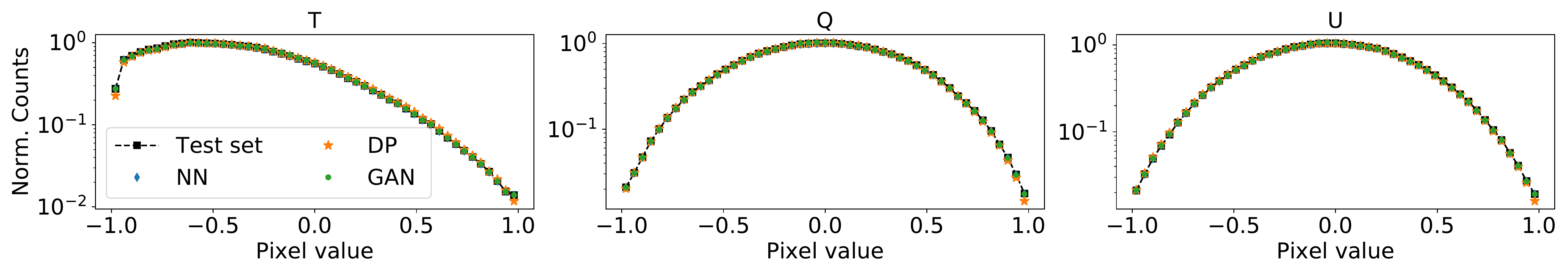}}
    { \includegraphics[width=2\columnwidth]{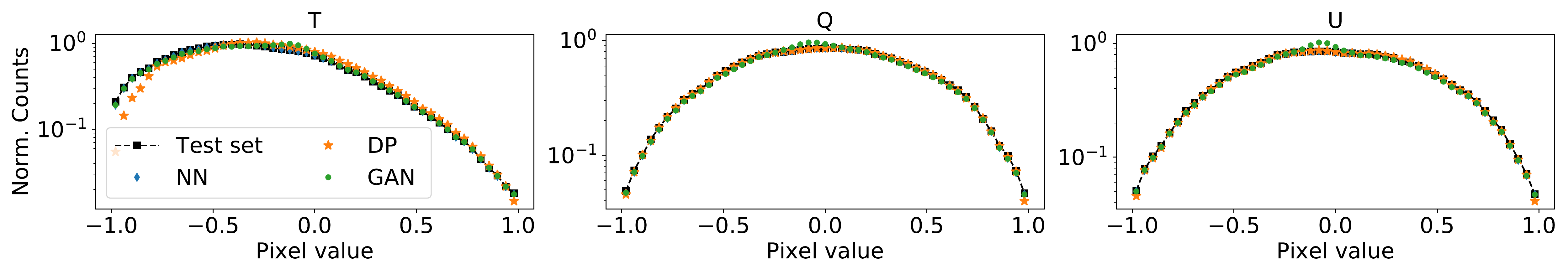}}
     \caption{Top: Thermal dust pixel intensity distribution of 500 generated maps with (diamonds) NN, (stars) DP,(circles) GAN compared to  500   maps from test set (black squares). Bottom: Synchrotron pixel intensity distribution.  The KS test statistics and the $p$-values listed in Tab.\ref{tab:pvals}  indicates that  the  distribution of inpainted images are likely to be drawn from the distribution of  the test set.  }
    \label{fig:low_order}
\end{figure*} 
\subsection{Evaluation of fidelity }\label{subsec:fidelity}

\noindent We employ several methods used in the literature to assess the quality of the reconstructed  maps quantitatively. First, we follow the approach from \citet{Mustafa17} and \citet{Aylor2019}, focusing on evaluating the ability to replicate the summary  statistics of the underlying signal that needs to be reconstructed. Those statistics are based on i) the pixel intensity distribution, ii) the angular power spectra of the two point correlation function, and iii) the first three Minkowski functionals.
We, therefore, consider the inpainting as \emph{successful} if it passes these three statistical tests.  
\begin{table}[htpb!]

    \begin{tabular}{c c c}
    \toprule 
  &  Synchrotron &   Thermal Dust      \\
  \midrule 
 NN      &$>0.997 $    &$>0.999$ \\ 
 DP &$> 0.963$&$>0.999$ \\
 GAN &  $>0.861$&$>0.997$  \\
 \bottomrule 
    \end{tabular}
    \caption{$p$-value of KS test performed on the pixel intensity distribution of Q maps shown in Fig.\ref{fig:low_order}. }
    \label{tab:pvals}
\end{table}

The distribution of pixel intensity provides information about whether the range of pixel values of the ground-truth maps are reproduced in the generated maps. Fig. \ref{fig:low_order}  shows the histogram of the pixel intensities of 500 generated maps compared with the corresponding ground-truth maps from the test set. For the types of foregrounds we consider in this analysis, the amplitude is strongly directional dependant. Therefore, we scale the sample maps with  \texttt{MinMax} rescaling to $[ -1,1 ] $ to reduce the patch-to-patch variation.  Although the differences  between the pixel distributions are nearly negligible, we run a Kolmogorov-Smirnov (KS) two-sample test on each case and assess how likely the distribution of inpainted images is drawn from the ground-truth distribution.  We thus estimate the empirical distribution function on the pixel samples from the ground-truth  images and from  images inpainted with three methods introduced in section \ref{sec:methods}, and then derive the KS two sample statistics to test the null hypothesis. From the KS \emph{p}-values summarized in Tab.\ref{tab:pvals}, we find that $p >0.86$ (0.997) for synchrotron (thermal dust) maps so that we can not reject the null hypothesis.

 \xedit{ In order to check whether  the Fourier modes in the ground-truth are reproduced in the inpainted maps,  we additionally evaluate the power spectra of intensity (TT), and polarization (EE and BB) maps}\footnote{Following the decomposition of Q and U maps  proposed in  \citet{Seljak_1997,Hu_1997}.} in each flat square map from the test set and the ones  inpainted with the three methods.

\begin{figure*}[htpb]
    \centering
  \includegraphics[width=2\columnwidth]{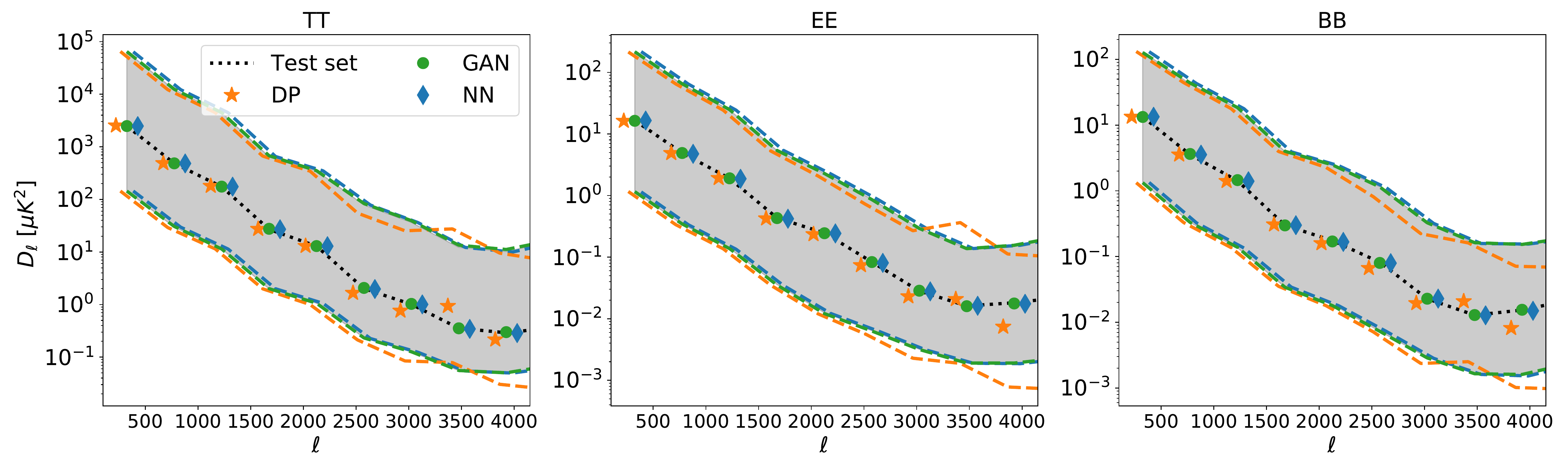}\\
  \includegraphics[width=2\columnwidth]{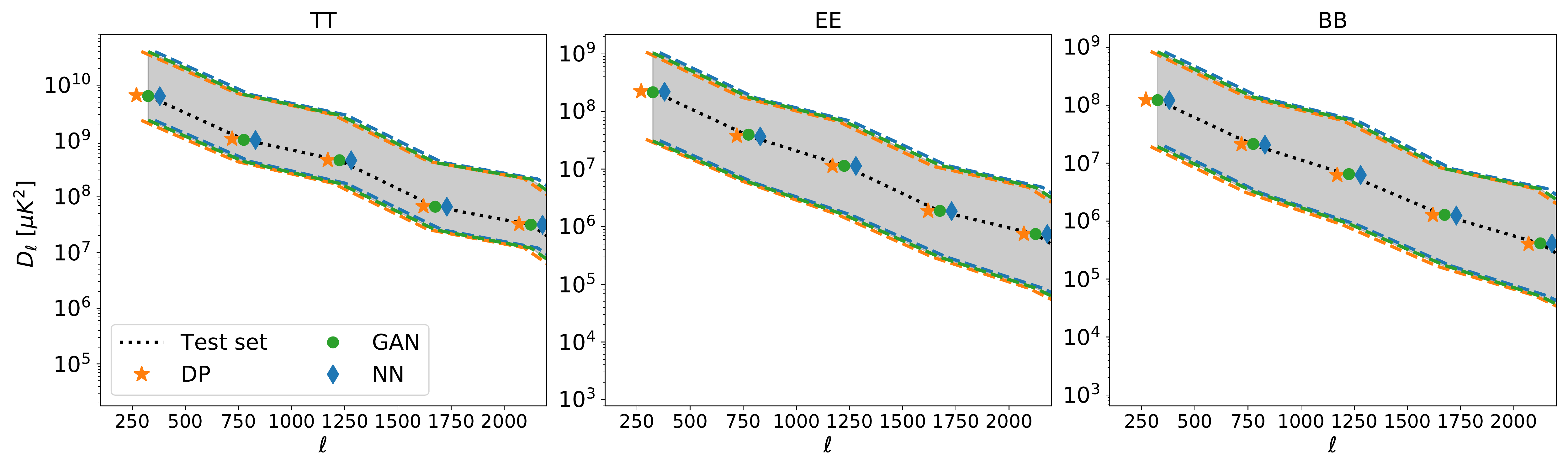}\\
     \caption{ From left to right,  TT, EE, BB power spectra estimated from a sample of  500  thermal dust (top) and synchrotron (bottom) maps.    Median power spectra are shown as (black dotted)  for the test set, (blue diamonds) for  NN  inpainting, (orange stars) for DP inpainting, (green circles) for GAN inpainting. The gray shaded area corresponds to the 95\% of test set  spectra to be compared with the 95 \% of the set of  power spectra  inpainted with NN, DP and GAN respectively shown as (dashed blue), (dashed orange) and (dashed green). The spectra are uniformly binned with $\Delta \ell =450$. }
    \label{fig:power_spec}
\end{figure*}

Each spectra shown  in Fig.~\ref{fig:power_spec}, is estimated with  \mbox{\textsc{Namaster}}\footnote{https://github.com/LSSTDESC/NaMaster} \citep{2019MNRAS.484.4127A}, binned  into  equally spaced multipoles with $\Delta \ell=450 $. The maximum multipole is chosen accordingly to the beam FWHM with whom  the signal is convolved, \ie $\ell_{max} = 4000 $ for dust and
 $\ell_{max} = 2000 $ for synchrotron. The median of the  binned power spectra is plotted at each multipole estimated from test set including 500 ground truth maps and corresponding inpainted maps.
 
The shaded-gray   area   represents 95$\%$ of the power spectra estimated from the test set  and its  vertical width  indicates how much the amplitude of the signal can vary at different locations of the sky (as much as 2 orders of magnitude).
The median power spectra estimated from  inpainting the test set are  shown  as points, and the area within the dashed lines corresponds to the 95$\%$ of the power spectra.   
  Notice that DP power spectra  for thermal dust  tend to systematically depart from the ones estimated with the  test set spectra  at $\ell >2500 $ scales. On the contrary, GAN and NN  are overall consistent with the    spectra from the ground-truth maps.

  To assess more quantitatively  that the power spectrum at a given $\ell$ bin is correctly reproduced, we consider  3 different multipole bins. We bootstrap resample 5000 times  the   distribution in each bin of 500 spectra and  perform the KS test on the resampled distribution.
  
\begin{figure*}
      \centering
      \includegraphics[width =2\columnwidth]{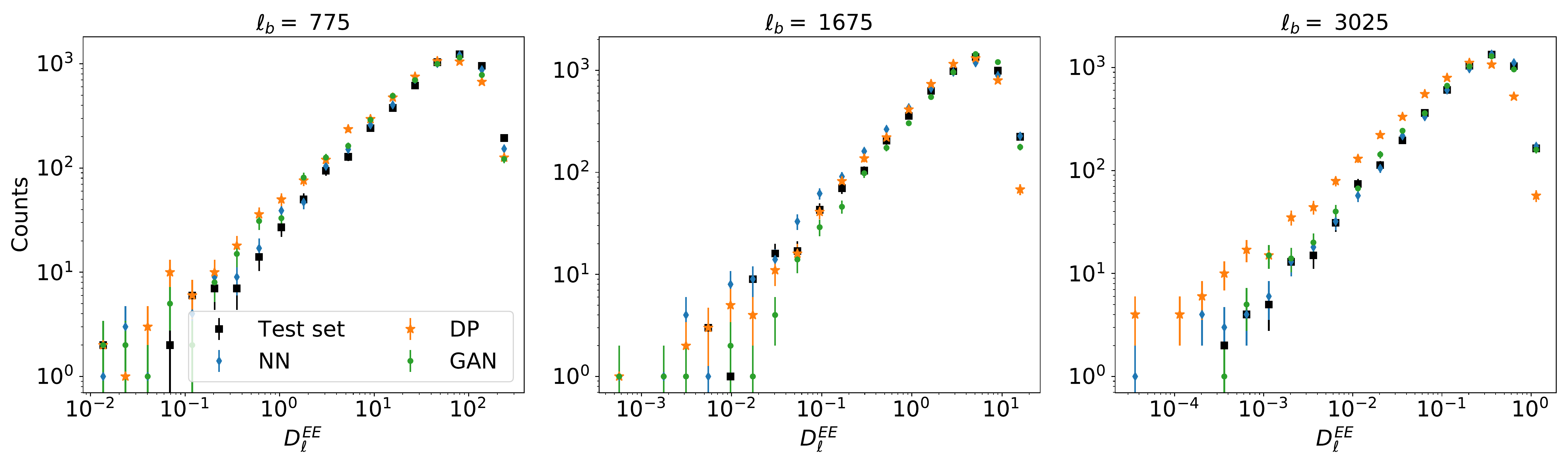}\\
      \includegraphics[width =2\columnwidth]{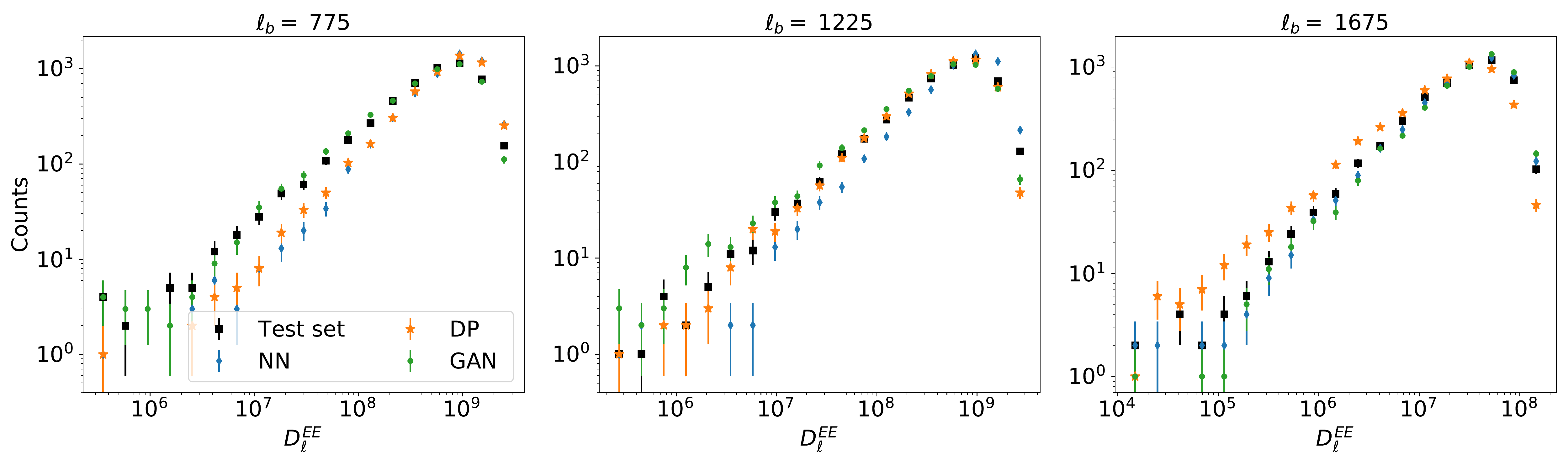}\\
      \caption{Distributions of EE power spectra of thermal dust( top panel) and synchrotron (bottom panel) at  3 different multipole bins  resampled   from the power spectra of 500 images inpainted with  (diamonds) NN, (stars) DP,(circles) GAN. We compare them with the resampled EE spectra   of   test maps (black squares).  { Error bars  are estimated as the squared root of number of elements in each bin after bootstrap resampling. } }
      \label{fig:bootstrap_spec}
  \end{figure*}

  Fig.~\ref{fig:bootstrap_spec} shows the distribution of EE spectra for thermal dust (top) and synchrotron (bottom) and favors GAN as the method that better resembles the ground-truth distribution, with  KS $p$-value  $>0.978$ $(>0.808) $ for dust (synchrotron) spectra.  On the other hand,   NN spectra present the lowest   $p$-values of $>0.808$ for dust and $>0.538$ for synchrotron.
 
\begin{table}[]

    \small  
    \begin{tabular}{c cc }
    \toprule 
  &  \multicolumn{1}{c}{Synchrotron} &\multicolumn{1}{c}{Thermal Dust }\\
  \midrule 
 \multirow{2}{*}{NN} &  $>0.306$ (2 bins)    & $>0.537$ (4 bins)     \\ 
                      & $>0.792$  (13 bins)  & $>0.801$ (23 bins)  \\ 
 %\midrule 
 \\ 
 \multirow{2}{*}{DP} &$>0.538 $ (5 bins)    & $>0.153$ (4 bins)    \\
                    &$  >0.792$  (10 bins) & $>0.788$ (23 bins)\\
\\ %\midrule 
 \multirow{2}{*}{GAN}&$>0.538 $ (5 bins)   & $>0.153$ (3 bins)   \\
                    &$  >0.792$  (10 bins)  & $>0.788$ (24  bins)  \\
 \bottomrule 
    \end{tabular} 
    \caption{$p$-value of KS test performed on  each multipole bin. We combined together  TT, EE, BB spectra, total 27 (15)  bins for dust (synchrotron).   }
    \label{tab:pvals_spec}
\end{table}
 
In Tab.\ref{tab:pvals_spec}, we summarize the KS test $p$-values estimated on each multipole bin and  after  having bootstrapped resampled the  TT, EE, BB spectra 5000 times in each bin. In total, we  account for ($3 \times 9$)  27 KS $p$-values for dust and ($3 \times 5$)  15 KS $p$-values for synchrotron spectra. Although there are few bins where the $p$-value is as low as $0.153$, the KS statistics  for those bins is small enough that the null hypothesis cannot be rejected at a significance level of $\alpha >  5 \% $. We therefore conclude that all the three  methodologies are able to reproduce angular correlations of the underlying signal coherently. 

\begin{figure*}[htpb]
    \centering
     \subfloat[  ] {
\includegraphics[width=2\columnwidth]{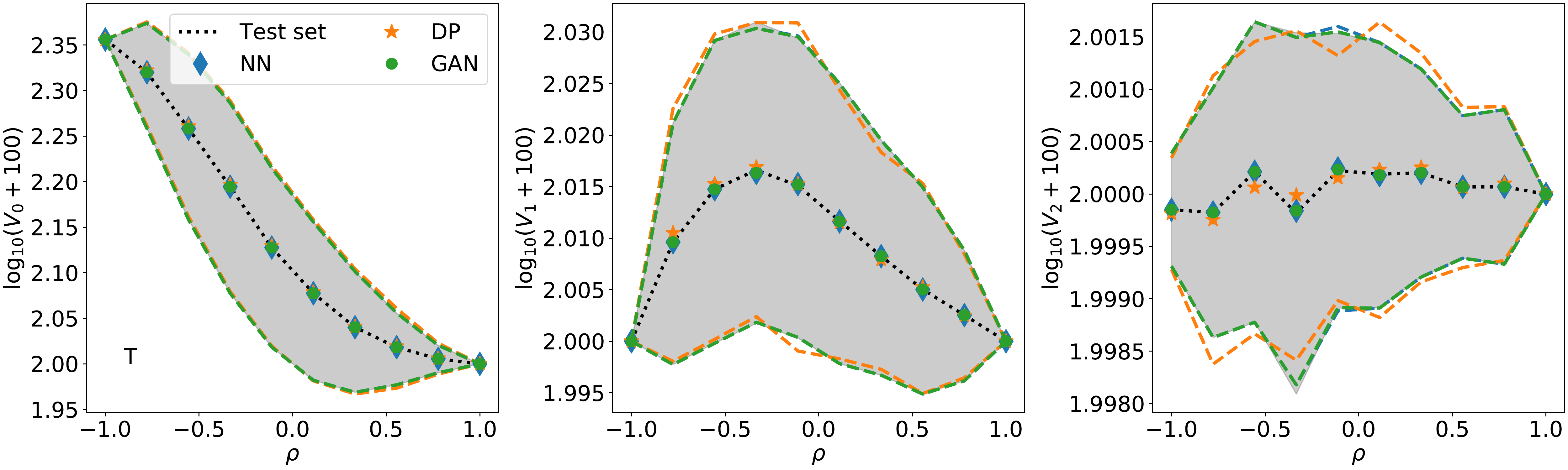} }

     \subfloat[] {\includegraphics[width=2\columnwidth]{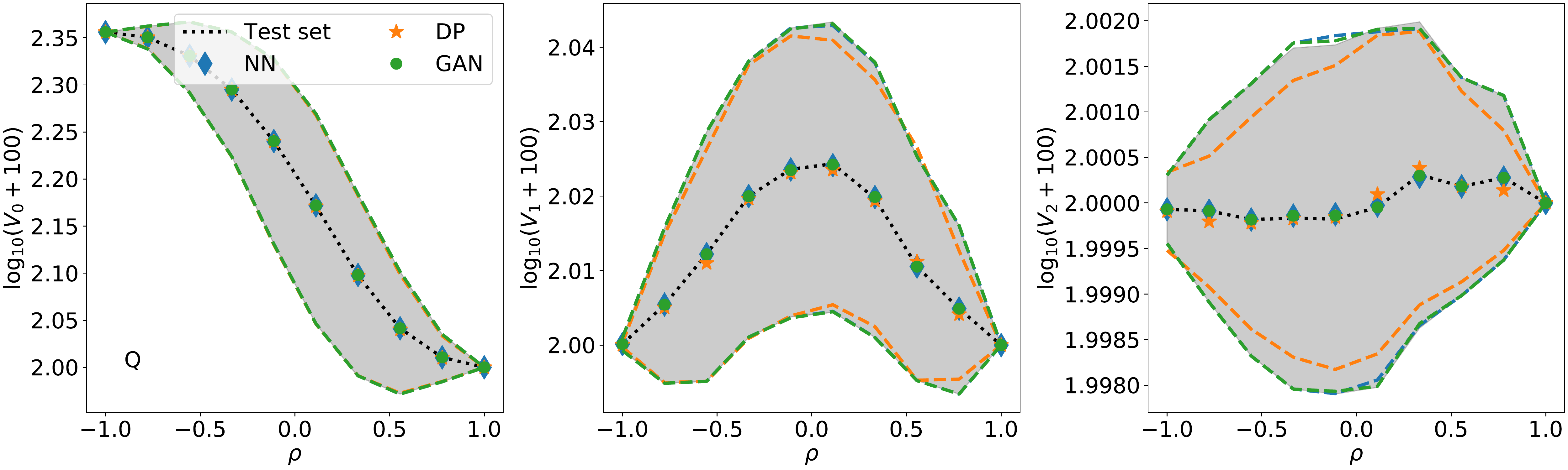}}

     \subfloat[] {\includegraphics[width=2\columnwidth]{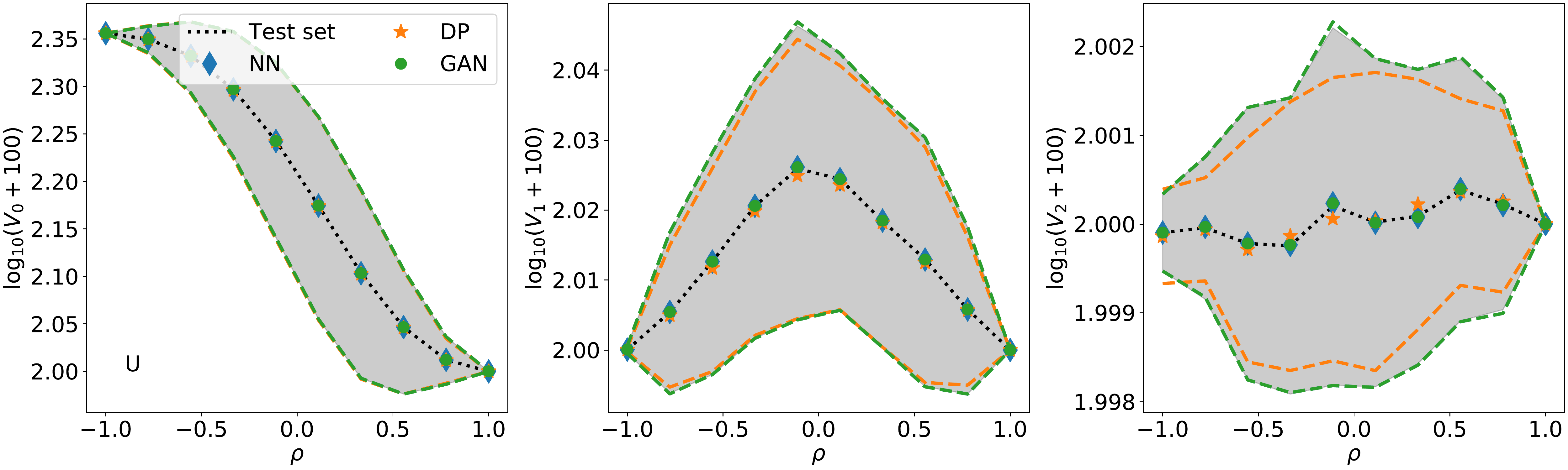}}
    \caption{ From left to right, $V_0,\, V_1,\, V_2 $ Minkowski functionals estimated from  the test set of  500 thermal dust T,Q and U maps respectively in (a), (b), (c). We use the same coloring scheme as in Fig. \ref{fig:power_spec}: (black dotted)  median of the functionals estimated from the test set, 95 \% of the functionals is shown as a gray shaded area . Points  and dashed lines refer to medians  and 95 \% interval of the functionals estimated from  the  sets of inpainted maps with (blue diamonds and blue dashed) NN, (orange stars and orange dashed)  DP and (green circles and green dashed) GAN.  }
    \label{fig:minko_dust}
\end{figure*}

\begin{figure*}[htpb]
    \centering
     \subfloat[  ]{\includegraphics[width=2\columnwidth]{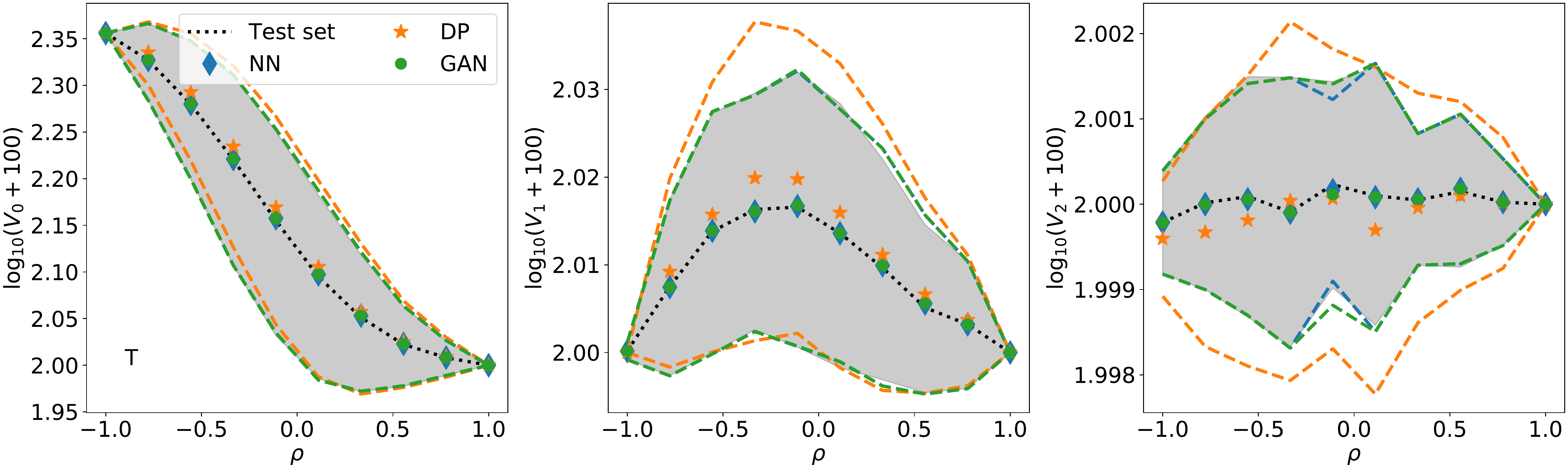}}

     \subfloat[  ]{\includegraphics[width=2\columnwidth]{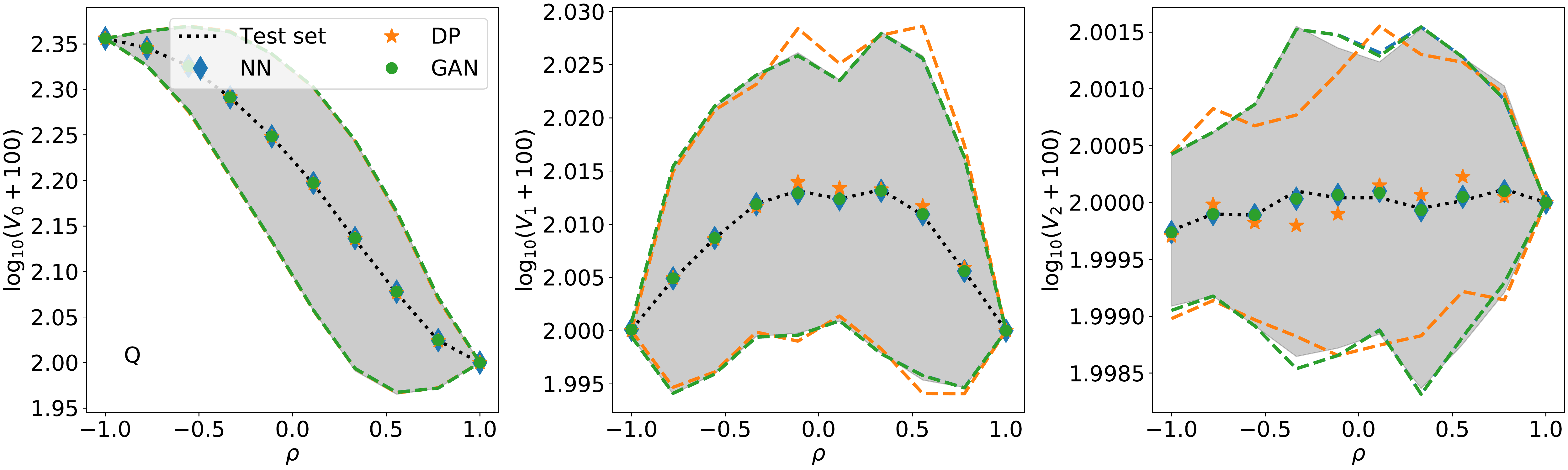}}
     
     \subfloat[  ]{\includegraphics[width=2\columnwidth]{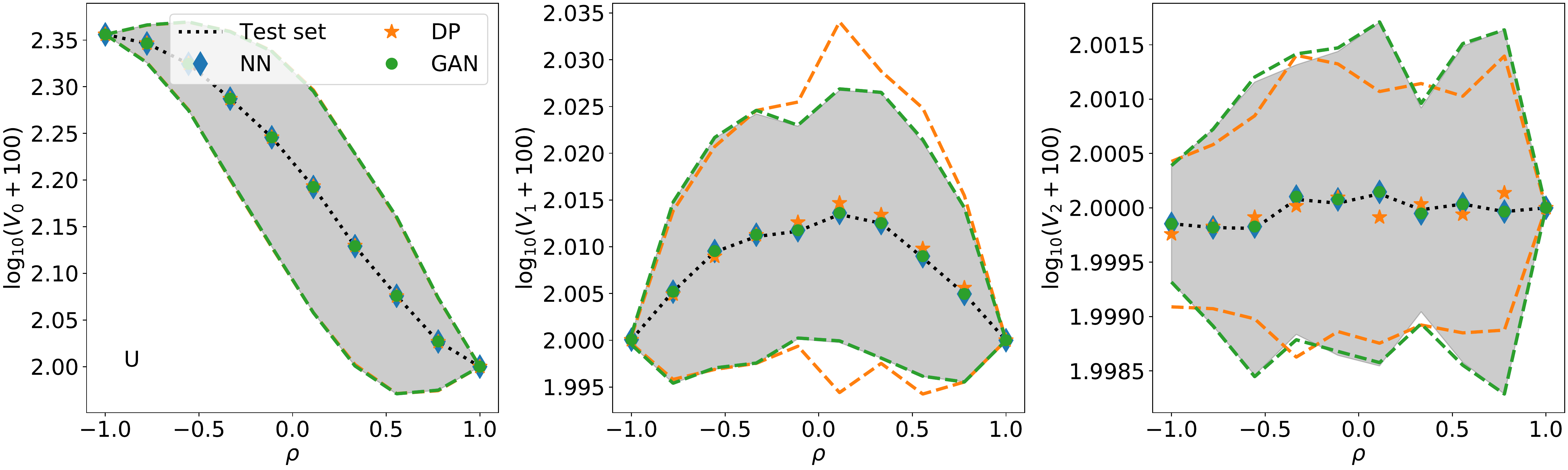}}
     \caption{ From left to right, $V_0,\, V_1,\, V_2 $ Minkowski functionals estimated from  the test set of  500 synchrotron  TQU maps respectively in (a), (b), (c).  We use the same coloring scheme as in Fig. \ref{fig:power_spec}: (black dotted)  median of the functionals estimated from the test set, 95 \% of the functionals is shown as a gray shaded area . Points  and dashed lines refer to medians  and 95 \% interval of the functionals estimated from  the  sets of inpainted maps with (blue diamonds and blue dashed) NN, (orange stars and orange dashed)  DP and (green circles and green dashed) GAN.    }
  
  \label{fig:minko_synch}   
\end{figure*}

Since the Galactic emission is highly non-Gaussian, it is {essential} to evaluate how well the methodologies described here are able to reproduce the non-Gaussian features. We thus evaluate the three Minkowski functionals $V_0, V_1, V_2$  for each  rescaled map (ranging in $[-1,1] $). The first three Minkowski functionals are related respectively to  the area, the perimeter and the connectivity in an image as a function of a threshold $\rho$. 
 The dotted black lines in Fig. \ref{fig:minko_dust} and \ref{fig:minko_synch}  show   the median   for the three Minkowski functionals (calculated using \citep{Mantz2008}) estimated from 500 samples and  evaluated at 10 equally spaced thresholds. Minkowski functionals for inpainted images are shown in  Fig. \ref{fig:minko_dust} and \ref{fig:minko_synch} with the same color scheme as in   Fig.\ref{fig:power_spec}. We notice that both GAN and NN are able to fully reproduce the non-Gaussianity of both the unpolarized and the polarized Galactic emission. 
 
 Similar to what is stated above, the Minkowski functionals estimated on the maps inpainted with DP clearly depart from the ground-truth especially at intermediate thresholds, ($-0.5<\rho<0.5$) for the $V_1$ and $V_2$ functionals. This can   point to further investigations for the DP inpaintings especially for better characterizing the non-Gaussian features and the small angular  scales $\ell > 2500$ of both dust and synchrotron that are not fully captured by the DP network.

\subsection{Validation on real data }\label{subsec:nulls}
\begin{figure*}[htpb!]
    \centering
    \subfloat[]{ 
    \begin{minipage}{2\columnwidth}

   \includegraphics[width=1.\columnwidth, trim=0cm 3.4cm 1.5cm 3.0cm , clip=true]{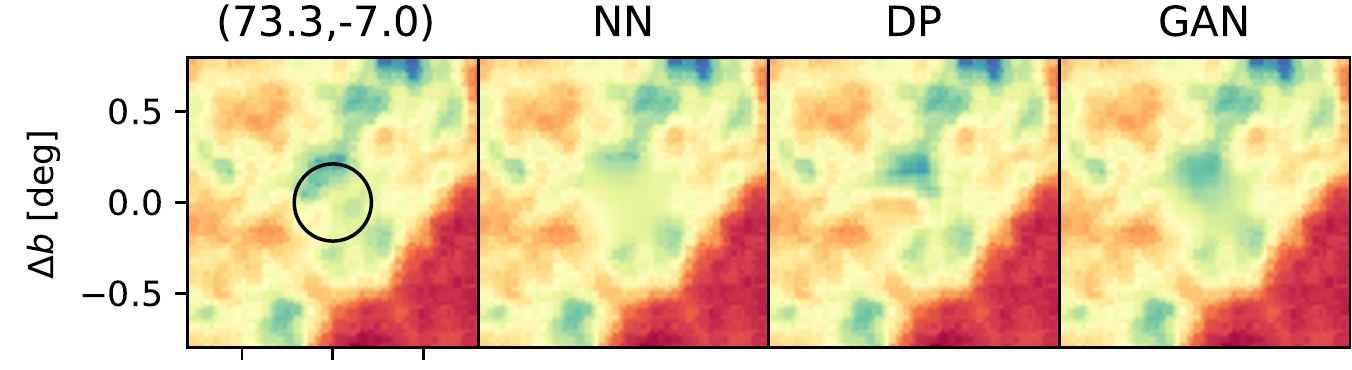}  
      \includegraphics[width=1.\columnwidth, trim=0cm 3.4cm 1.5cm 3.0cm , clip=true]{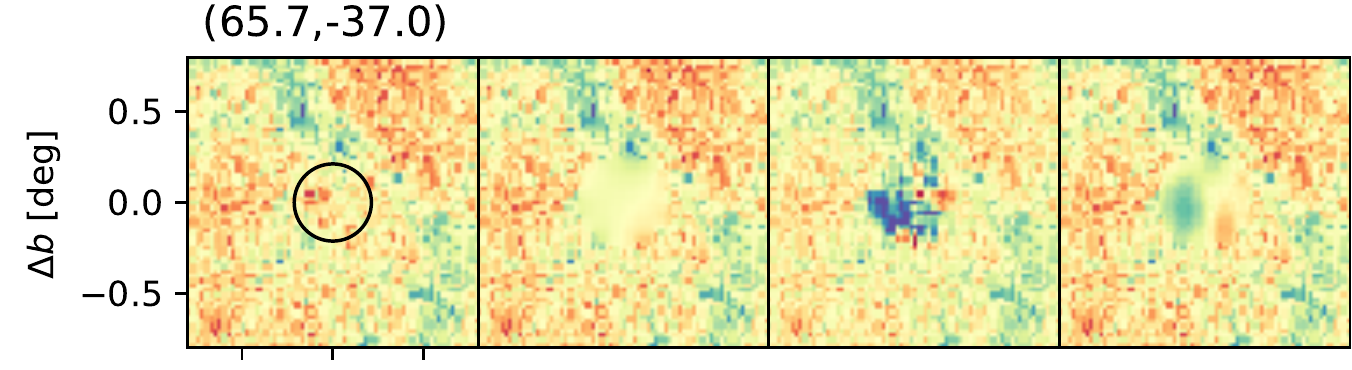}
      \end{minipage}
      }

    \subfloat[]{ 
    \begin{minipage}{2\columnwidth}
   \includegraphics[width=1.\columnwidth, trim=0cm 3.4cm 1.5cm 3.0cm , clip=true]{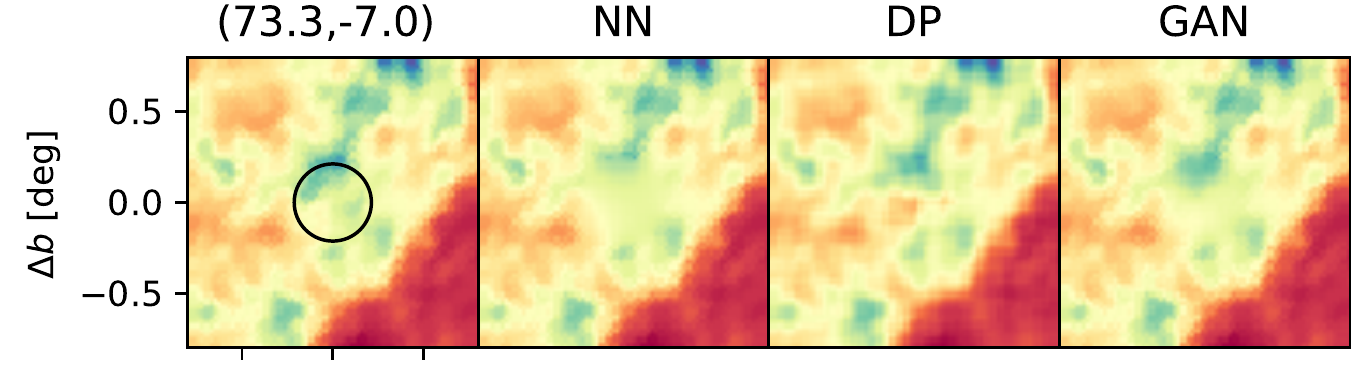}\\

   \includegraphics[width=1.\columnwidth, trim=0cm 2.5cm 1.5cm 3.0cm , clip=true]{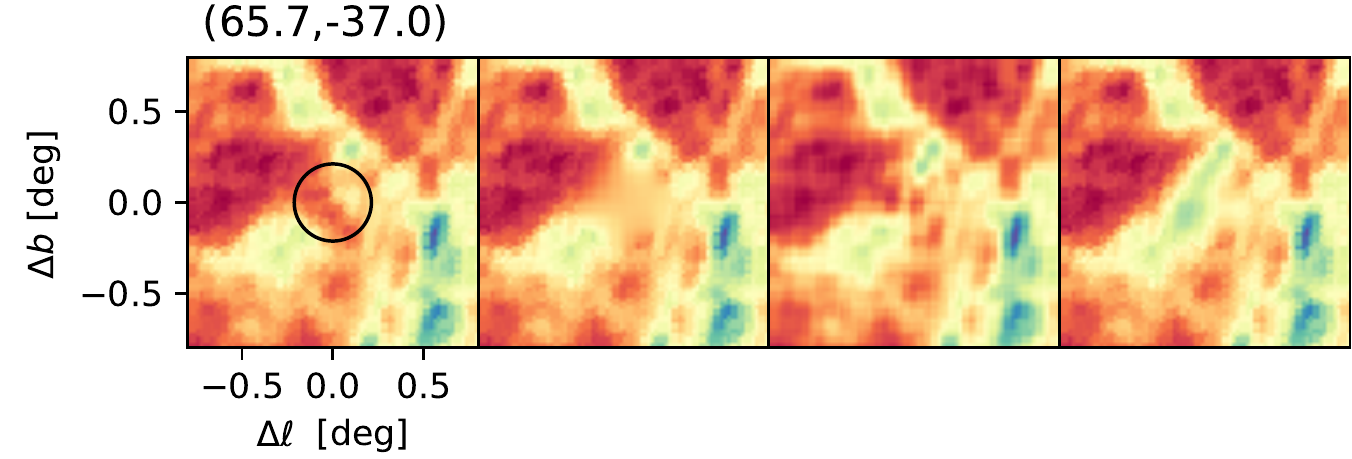}\\
   \end{minipage}
 }    
 \caption{ Thumbnail images  of \emph{Planck} temperature    maps at (a) 353 and (b)  857 GHz. The radius of the  reconstructed area (black circle) is 15 \si{\arcmin}. Two  locations  are chosen to highlight different inpainting  performances    with high and low SNRs of  \emph{Planck} 353 map, respectively at high Galactic latitude, \ie   $(l,b) = (65.7\si{\degree}, -7\si{\degree}) $ and at low Galactic latitude,\ie  $(l,b) = (73.3\si{\degree}, -37\si{\degree}) $.}
    \label{fig:null_inpainting}
\end{figure*}
\noindent To further test and validate our methodologies, we run tests on real data, cropping 500 images at random locations from the \emph{Planck} maps at 353 and 857 \si{\giga \Hz}.

\begin{figure*}[htpb!]
\centering 
\subfloat[]  { 
\begin{minipage}{2\columnwidth}

\includegraphics[width=.5\columnwidth]{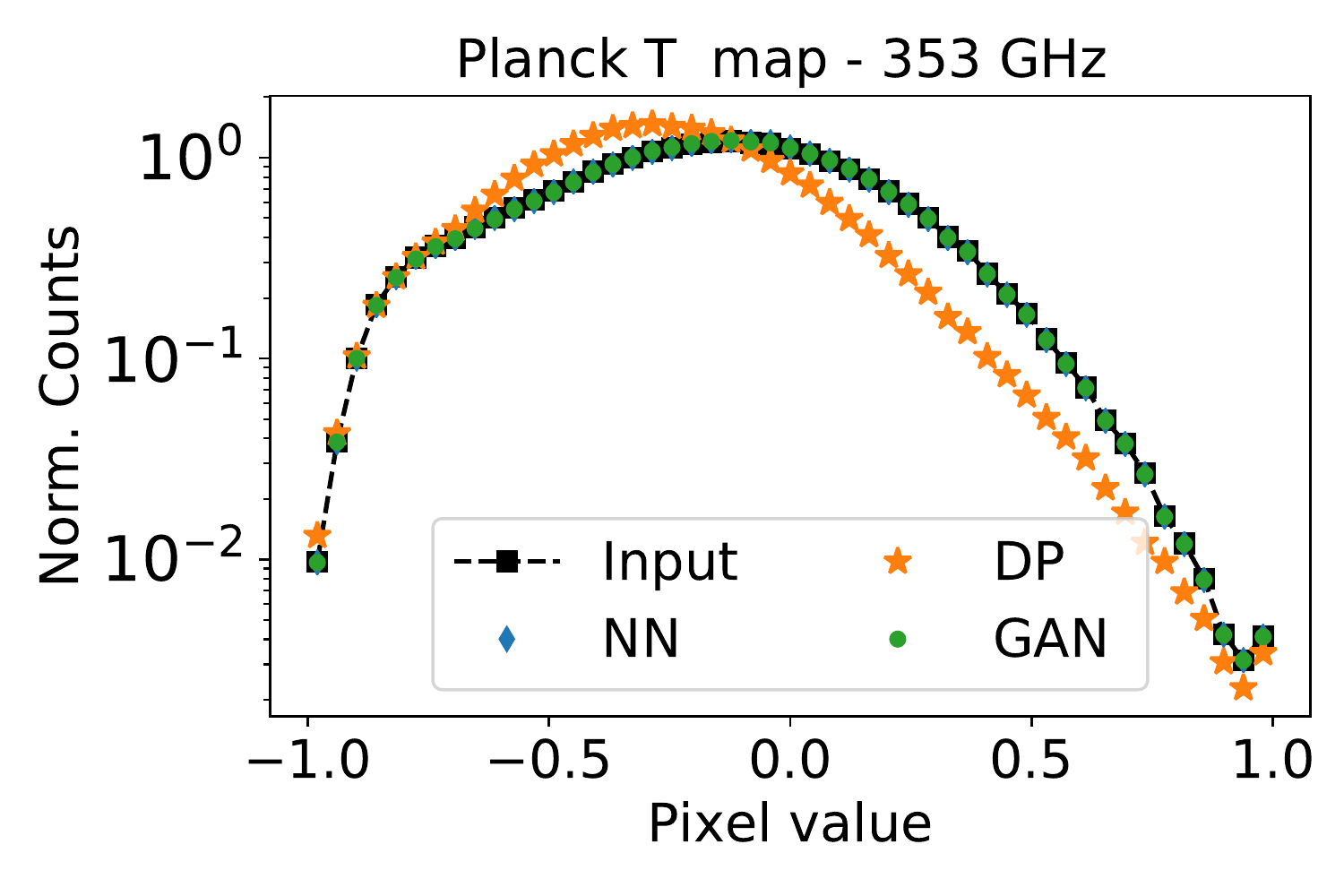}
\includegraphics[width=.5\columnwidth]{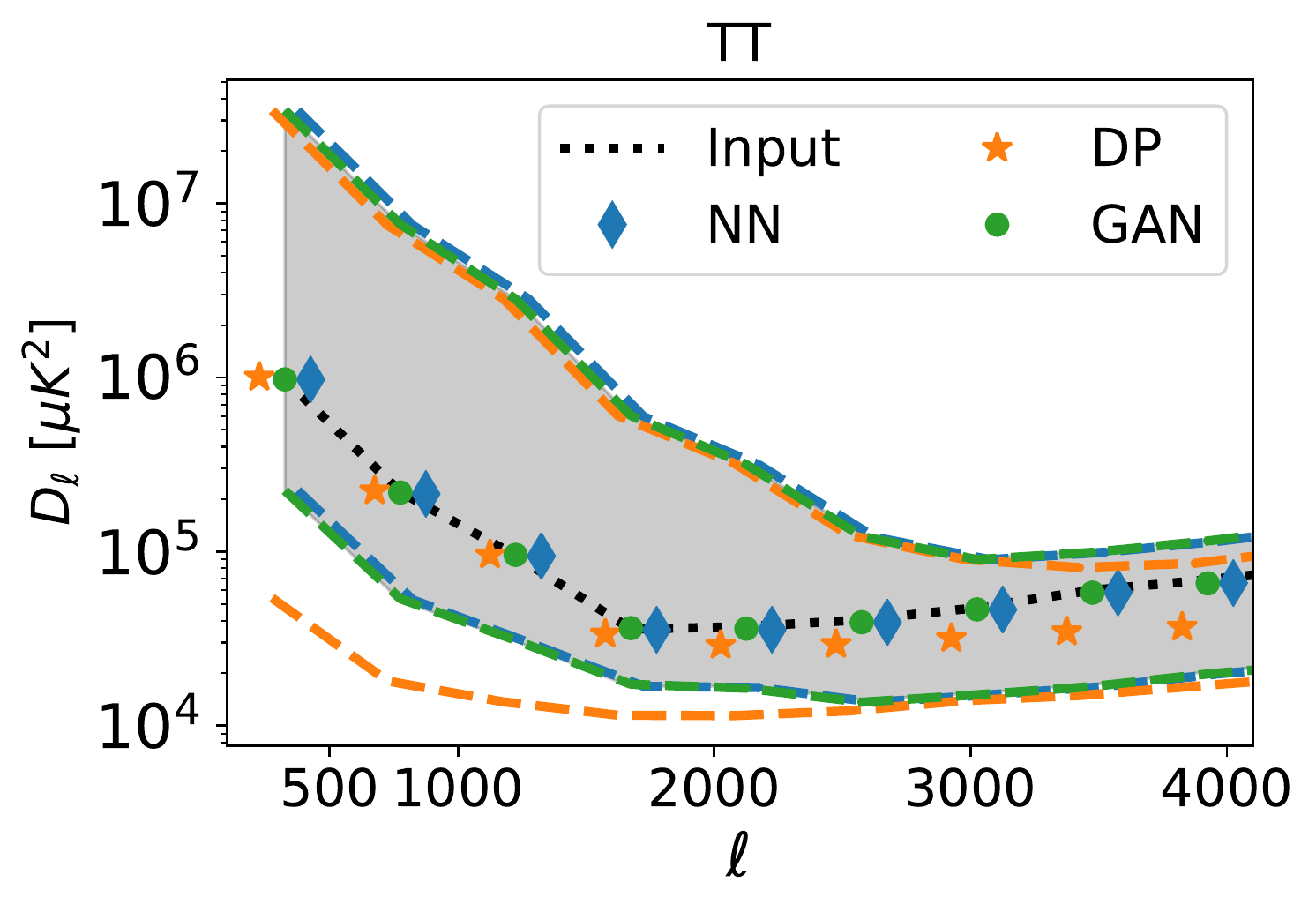} 
\includegraphics[width =1.\columnwidth ]{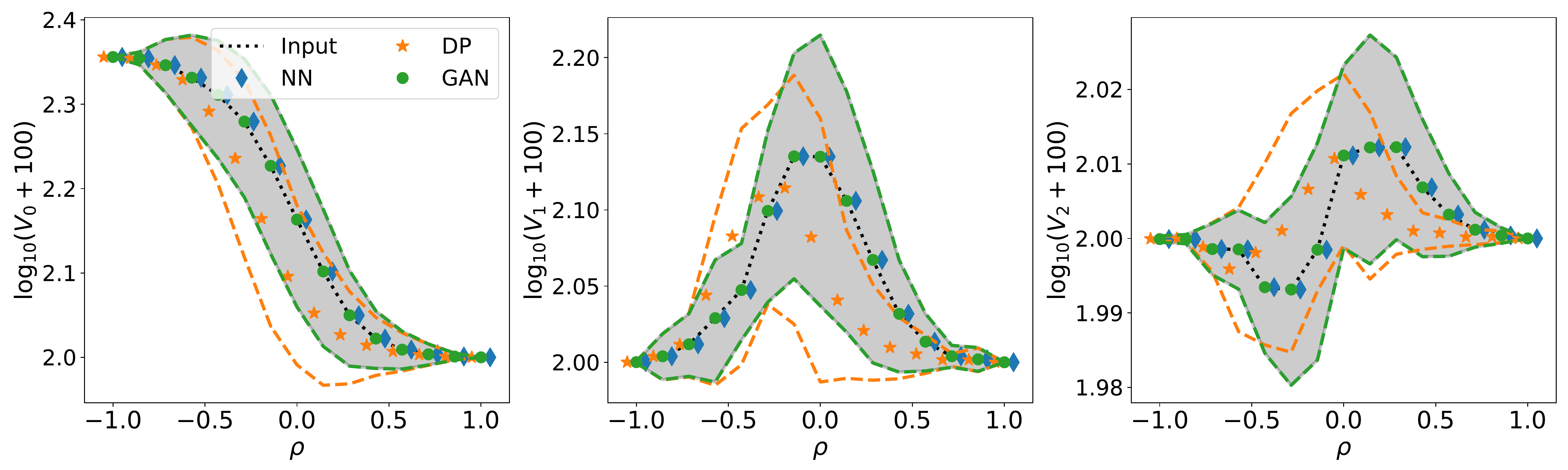}
\end{minipage}
} 
\\
\subfloat[]  { 
\begin{minipage}{2\columnwidth}
 \includegraphics[width=.5\columnwidth,  ]{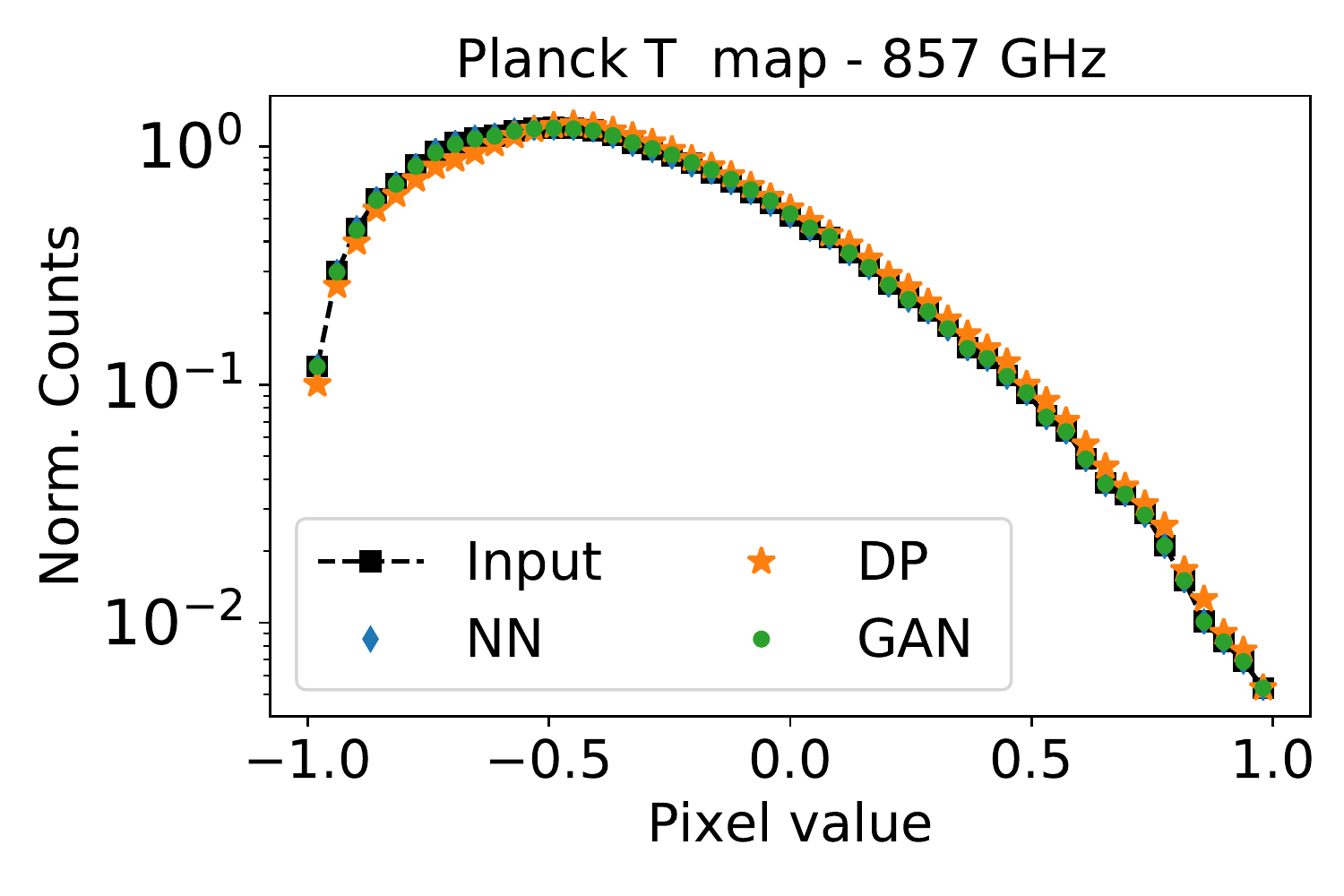}
\includegraphics[width=.5\columnwidth]{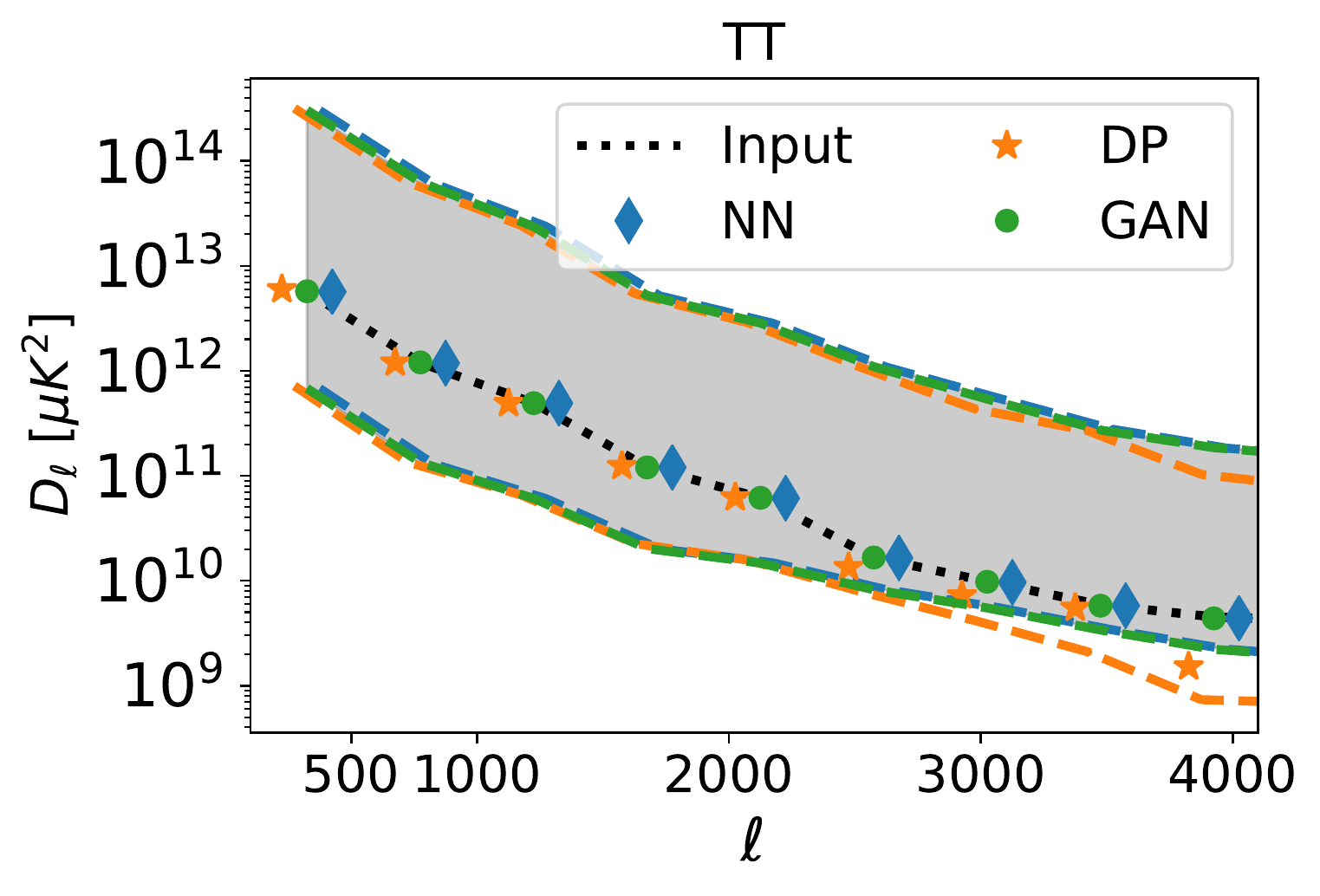}

\includegraphics[width =1\columnwidth ]{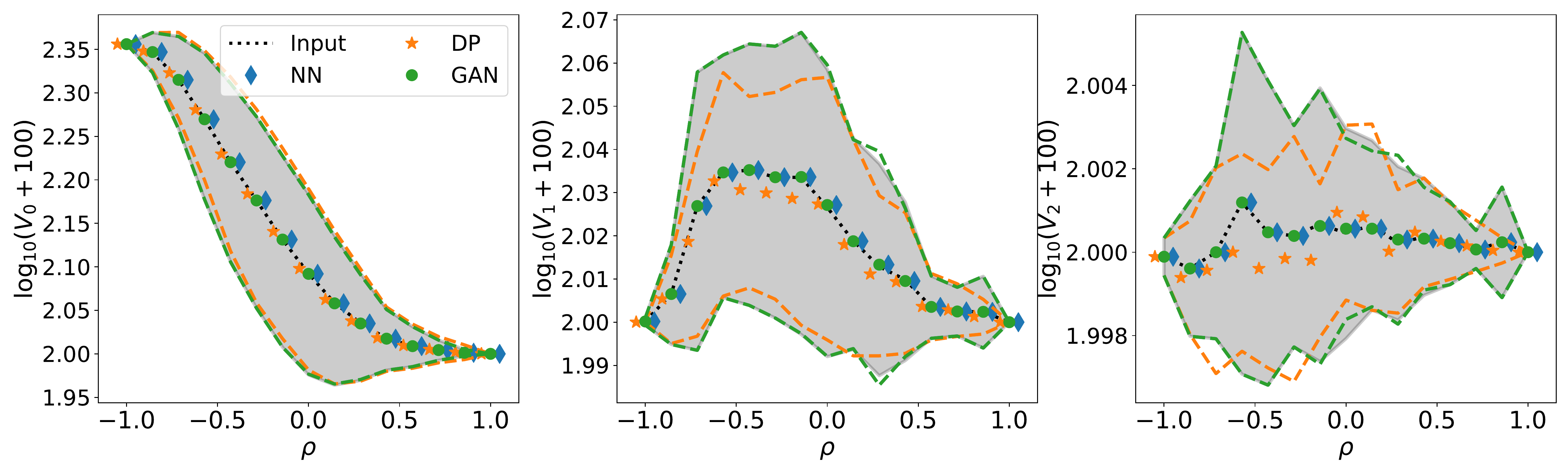}
\end{minipage} 
} 
  \caption {Summary statistics estimated on \emph{Planck} dust T  maps at (a) 353 and (b) 857 \si{\giga \Hz}.}   \label{fig:null_summary} 
\end{figure*}

\begin{figure*}[htpb!]
\subfloat[   ]{
\begin{minipage}{2\columnwidth}
\includegraphics[width=1\columnwidth, trim=0cm 3.4cm 1.5cm 3.cm , clip=true ]{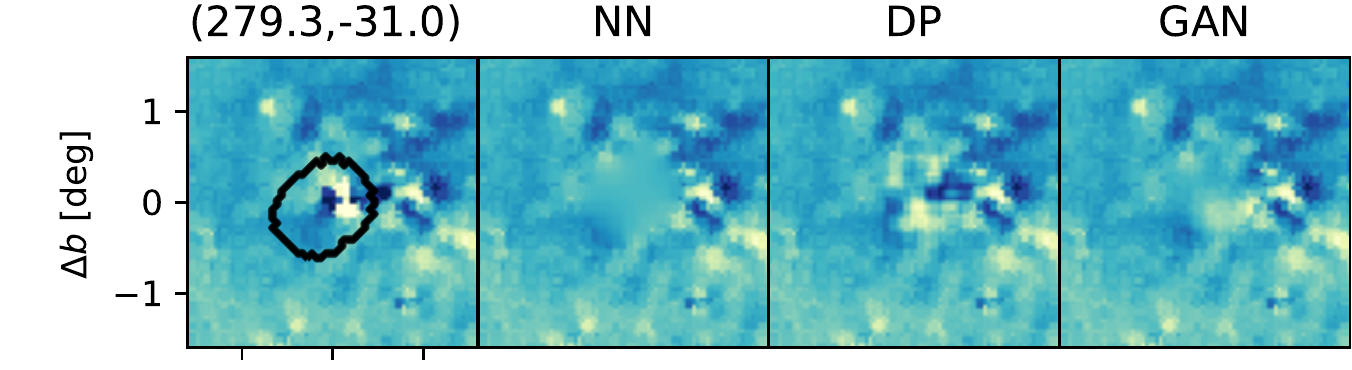}
  \includegraphics[width=1\columnwidth, trim=0cm 3.4cm 1.5cm 3.cm , clip=true ]{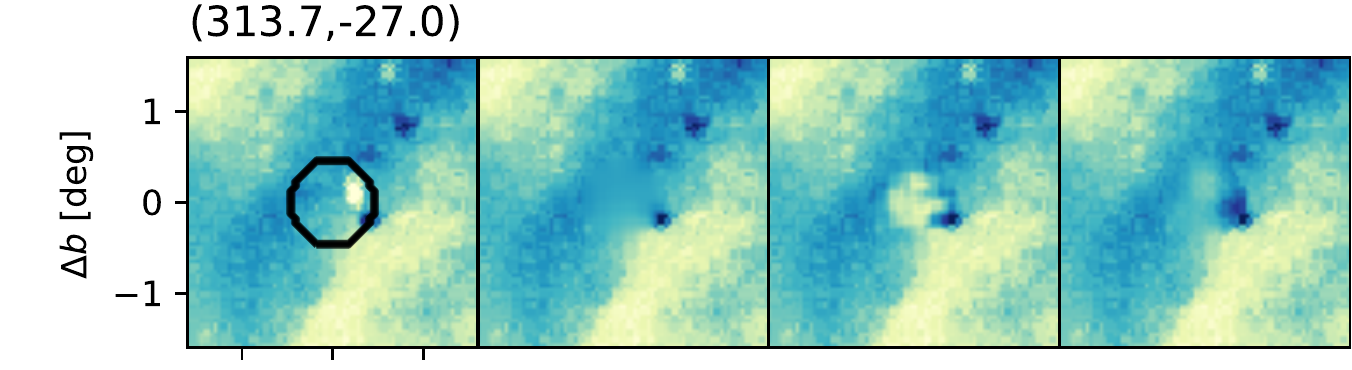}
  
  \end{minipage}
  }
  
 \subfloat[     ]{
\begin{minipage}{2\columnwidth}
  \includegraphics[width=1\columnwidth, trim=0cm 3.4cm 1.5cm 3.cm , clip=true]{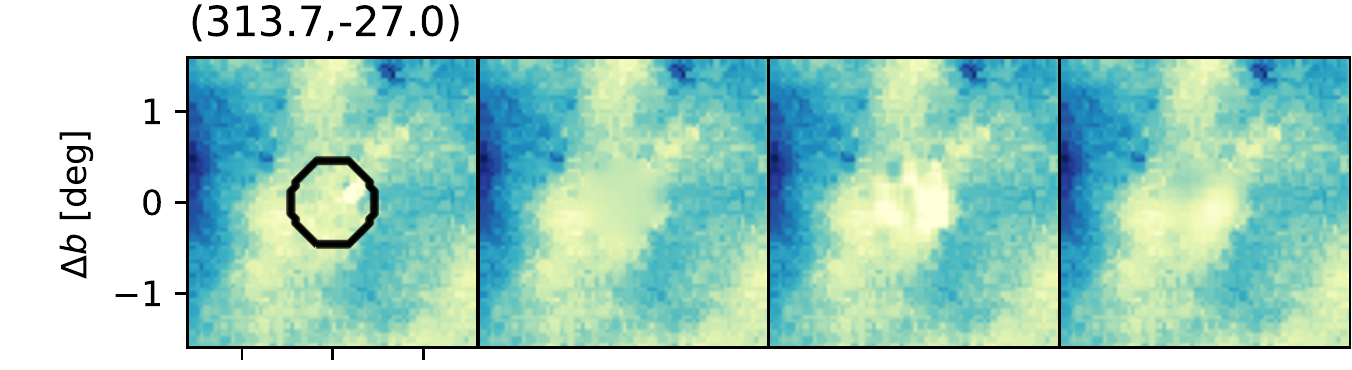}
  \includegraphics[width=1\columnwidth,trim=0cm 3.cm 1.5cm 3.cm , clip=true]{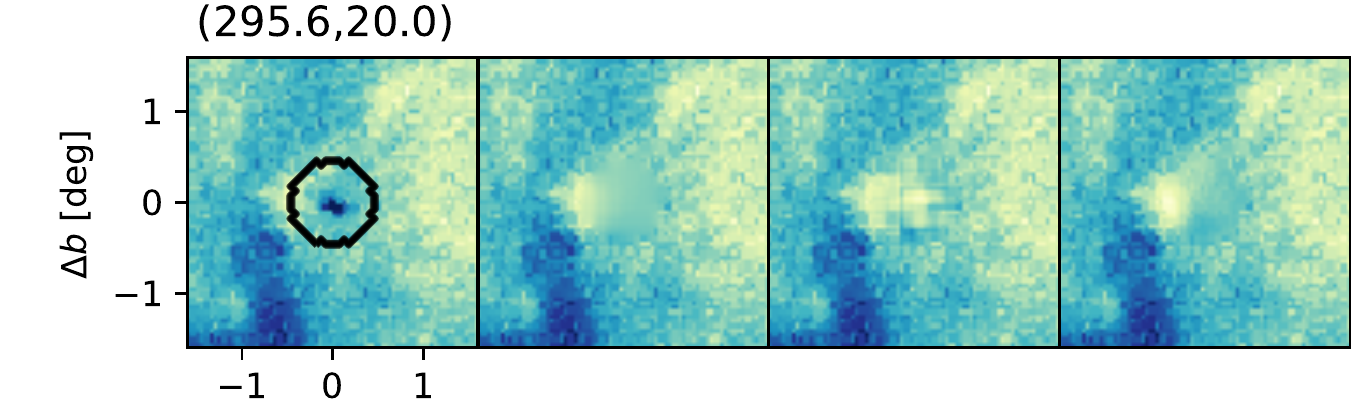}
  \end{minipage}
  }

    \caption{From left to right,  Input SPASS, DP, NN and GAN polarization maps inpainted in 30 arcmin region surrounding the detected  point-source (black circle). The range of the input map is chosen to be the same as the one of the inpainted ones.  Temperature maps are shown in  Fig.\ref{fig:inpaint_Tspass}.}
    \label{fig:inpaint_spass}
\end{figure*}

Fig.\ref{fig:null_inpainting} shows the $1.5 \times 1.5 \,  \si{\deg} ^2$  \emph{Planck} maps  at (a) 353 and (b) 857 GHz. Notice that we deliberately choose two locations to highlight reconstruction performances at two different SNR regimes.  Closer to the Galactic midplane, the  dust emission is stronger so that SNR is  high at both 353 and $857\, \si{\giga\hertz}$, \eg  see top  panels of Fig.\ref{fig:null_inpainting}(a) and (b). On the contrary, at high Galactic latitudes and at 353 \si{\giga\hertz}, the SNR can be lower and noise contribution is clearly noticeable in the maps, \eg as in the bottom panel of Fig.\ref{fig:null_inpainting}(a).  { In this case,  the inpainting  performances with DP and GAN can be   affected by noise, and reconstruction area can be visually distinguished in the square patch.}

For a more quantitative assessment, we thus estimate the summary statistics shown in Subsect.\ref{subsec:fidelity} and we show the results in Fig.\ref{fig:null_summary}. 
 
A clear indication that the thermal dust emission data is contaminated by instrumental noise can be inferred by comparing
   the shapes of  Minkowski functionals estimated for  the maps  at 353 \si{\giga\hertz}, with the ones    from  signal-only simulations  (fig.\ref{fig:inpaint_dust}) or  from  signal-dominated maps (fig.\ref{fig:null_summary} (b)). The morphology of the Minkowski functionals of the former  largely  resembles the functionals estimated from  a Gaussian signal.  Two possible candidates for the Gaussian component are: i) \emph{Planck} instrumental noise which   can be approximated as  white within   $\sim 9  \, \si{\deg}^2$   patches  and ii) CMB residual emission.
    
  DP inpaintings are the most affected ones in the presence of the noise at $353\,  \si{\giga\hertz}$,  \eg notice how the pixel distribution (top left panel) of Fig.\ref{fig:null_summary}(a) noticeably departs    from the ground truth one. 
However, the KS test $p$-value performed on the DP and ground truth  samples is $p> 0.536 $, \ie  not significantly low enough to reject the null hypothesis that the two samples are different. 
On the other hand, inpainting with GAN and NN does not show any dependence with SNR, as the performances with these methodologies  are essentially similar to the ones observed with the signal-only simulated maps (\eg Fig.\ref{fig:low_order}, \ref{fig:power_spec} and  \ref{fig:minko_dust}). 

Finally, we would like to point out that for the case of inpainting with  GAN, we use the weights derived  from the training set that composed of  signal-only dust TQU simulated maps.  Looking at Fig.\ref{fig:null_summary} we notice that GAN is able to statistically reproduce at 353 GHz  the features composed by signal and noise, indicating that the network has correctly learned the features related to the intrinsic signal in presence of noise, and injects signal plus noise features statistically coherent with the ones outside the masked area. 
However, when the noise is highly dominating in the patch, we can  clearly distinguish smooth artifacts in some cases inpainted with GAN (see  Fig.\ref{fig:null_inpainting}(a, bottom)). This is somewhat expected because  GAN is trained on signal-only simulated images. Further investigations on training GAN with noisy data are needed, and we will address it in a future work.

\begin{figure*}[ htpb!]

\begin{minipage}{2\columnwidth}
\includegraphics[width=1\columnwidth, trim=0cm 3.4cm 1.5cm 3.cm , clip=true ]{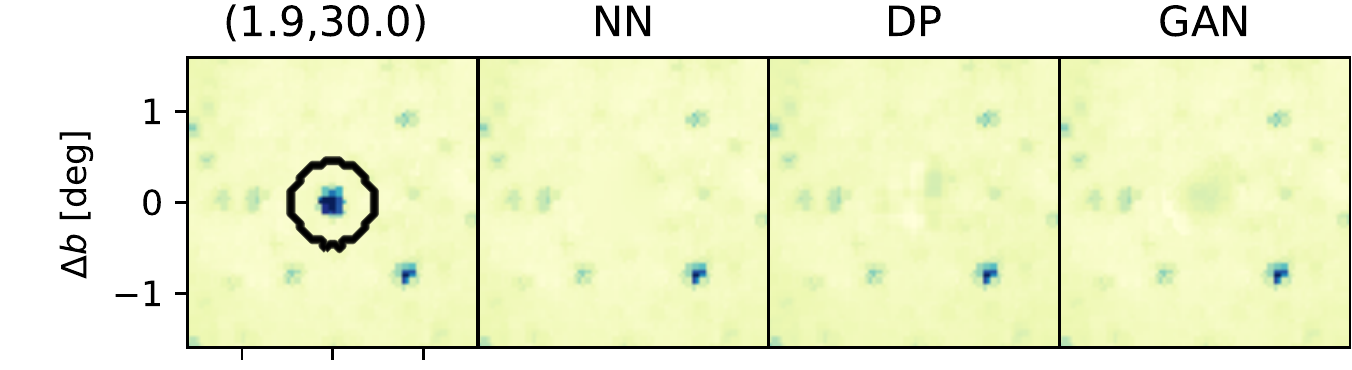}
\includegraphics[width=1\columnwidth,trim=0cm 2.5cm 1.5cm 3.cm , clip=true]{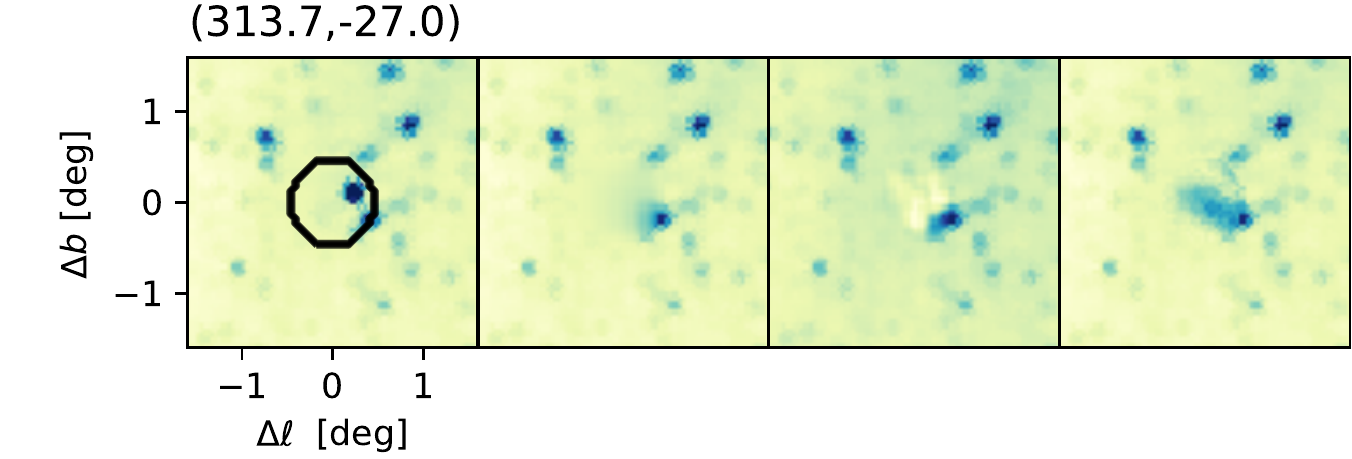} 
  \end{minipage}

  \caption{  SPASS temperature  maps inpainted in 30 arcmin region surrounding the detected  point-source (black circle). The range of the input map is chosen to be the same as the one of the inpainted ones.  }
    \label{fig:inpaint_Tspass}
\end{figure*}

 \subsection{Inpainting maps with Point-Sources }\label{subsec:spass}

\noindent   In this section, we aim at showing real world applications of the three reconstruction methodologies by inpainting areas in real maps with extra-galactic radio sources. 

  We consider the TQU SPASS synchrotron map at $2.3 $ \si{\giga \Hz}\footnote{Maps are  available online at \url{https://sites.google.com/inaf.it/spass}.  } ,and we run a \emph{Matched Filter }\citep{Marriage2011} to detect the brightest polarized sources in the map. We consider $7 \sigma$ as a  threshold for a point-source detection. In particular, we focus on unresolved point-sources (\ie sources whose projected solid angle is smaller than theSPASS beam solid angle) detected at intermediate  Galactic latitudes $|b|>20 $ \si{\deg}. As a result, 45  polarized sources are detected in the SPASS map, which is very close to 60, the number forecasted  by  PS4C with the adopted SPASS specifications.

   Fig.\ref{fig:inpaint_spass} shows a selection of images extracted from the SPASS TQU maps and centered at  the coordinates of the detected sources. The shape and the size of the region to be inpainted are chosen proportionally to the flux of each source (see the black circle in fig.\ref{fig:inpaint_spass}). However, we expect the inpainting not to be affected by different shapes and/or sizes of the masked area as demonstrated in  \citet{generative} and \citet{deeprior}. Moreover, we set the color scale of the input SPASS map to be the same as the one in  the inpainted maps to highlight the consistency with reconstructed maps.  As expected, images inpainted with GAN are visually injecting more coherent features and less artifacts with respect to DP and NN.  
   
   Given the presence of the point-source at the center of the patch biasing the evaluation of fidelity with the pixel distribution and the Minkowski functionals, we estimate the power spectra from all the sets of maps in order to assess more quantitatively the quality of the generated maps. We masked the source with a circular mask with 30 \si{\arcmin} radius and estimate the spectra  in  the  area outside  the mask. On the other hand, we did not apply the point-source mask for the power spectra estimated from the inpainted maps. 
      \begin{figure*}[htpb]
\centering 
\includegraphics[width=2\columnwidth, trim=0cm 0cm 0cm 0cm , clip=true]{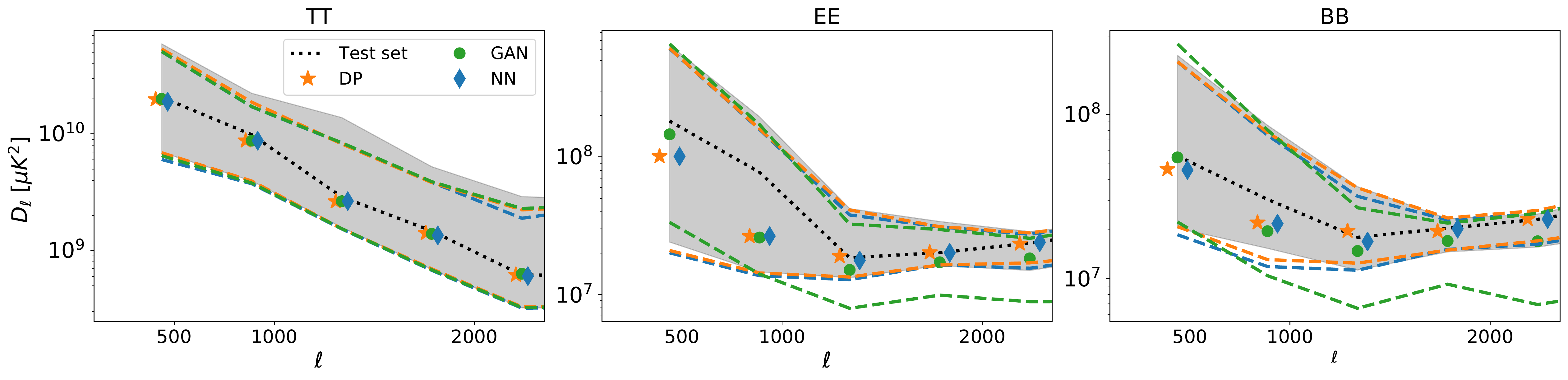}
\caption{ TT, EE, BB power spectra estimated on the SPASS map. In this case,  the  input map  encodes a point-source located at the center and  is masked out  before computing the power spectra. Vice versa, spectra estimated on inpainted maps do not have the mask applied. }\label{fig:spass_spectra}
\end{figure*}

   We estimate the power spectra as described in the previous sections. In Fig.\ref{fig:spass_spectra}, we show the TT, EE and BB power spectra, which indicates all  methodologies essentially are able  to reproduce consistently the power spectrum at all angular scales of the input SPASS masked maps. Moreover, on smaller angular scales, the power spectra from maps inpainted with GAN show a lower amplitude tail, possibly implying that Poissonian bias from undetected point-sources is further reduced (a factor of $\sim 4 $ for EE and BB spectra).

\section{Code  Release }
\noindent The three inpainting methods have been collected into  a \texttt{python} package, the Python Inpainter for Cosmological and AStrophysical SOurces (\textsc{Picasso} \href{ github.com/giuspugl/picasso}{\faGithub}).  It  has  been  made publicly available together with a documentation web page\footnote{\url{http://giuspugl.github.io/picasso/}}.  

 We trained GAN separately on dust and synchrotron map sets\footnote{Training weights can be downloaded from \url{https://bit.ly/2TI6x4o} }. The training process for each set took $\sim 12 $  hours on a GPU node of \textsc{Sherlock} Cluster of Stanford Supercomputing Center with 4 interconnected  NVIDIA Tesla P40 GPUs \footnote{For more details,  see \url{https://www.sherlock.stanford.edu}. }.

Finally, we measure the inpainting speed  for each of the three methods. We  performed this benchmark within a  GPU node at NERSC equipped with  4 interconnected NVIDIA Tesla -V100 GPUs\footnote{\url{https://docs-dev.nersc.gov/cgpu}}.  GAN is the fastest method since   fast-forwarding the trained weights is very  quick and it takes $\sim4\div6$ \si{\second}.  Although NN does not involve any DCNN in the map reconstruction, it iterates over the pixels in the missing region and it takes $ \sim 15 $ \si{\second}  per image. A single inpainting with  DP  takes $\sim30$ \si{\second}, since this is the time spent to  minimize the loss function in eq.\ref{eq:lossDP} over 3000 - 5000 epochs with gradient descent.  

\section{Discussion}

 {  
\noindent In the near future,  several high-resolution experiments from the ground, \eg Simons Observatory, the South Pole Observatory and CMB-Stage IV\citep{2019JCAP...02..056A,2019BAAS...51g.209C,2019cmbs4}, are expected   to detect more than ten thousands sources in total density flux   representing  a critical contaminant especially at small angular scales \citep{N_ss_2019, Lagache2019,puglisi2018}. This is mostly  due  to an improvement in  sensitivity  and an increase in the  footprint  area ($\sim 40 \%$ of the sky)  with respect to the ones that have been surveyed so far.} 

 { To mitigate the contamination issue, detected point sources are usually masked out from the maps. However, a mask encoding more than 10,000 can result in biasing the 2-point (as well as the higher-order) correlation functions used to estimate cosmological parameters. } { Several methodologies have been proposed in the literature to clean the maps from  point sources: e.g. if the source is unresolved by fitting the beam profile at the map level \citet{N_ss_2019,2019MNRAS.486.5239D} as well as  inpainting  with several methodologies, not necessarily involving DCNNs \citet{Bucher_2012}. } 

 { On the other hand, point sources  contaminate the Galactic foreground emission especially far from the Galactic mid-plane (\ie at high Galactic latitudes) . A mitigation of point source contamination in polarized foregrounds maps at small angular scales is thus  needed in order to have \emph{de-lens} the gravitational lensing B-mode and assess the amplitude of the  primordial B-mode.}

In this work, we inpaint thermal-dust and synchrotron   intensity and polarized emissions in order to deliver point-source-free foreground template maps. We show that not only the pixel distribution and two-point correlation function of the inpainted maps, but also their higher statistic moments commonly adopted for non-Gaussian maps, \ie  the Minkowski functionals, are  statistically  consistent  with the ones from the real data set by means of the KS test. Given  the  fact that  applications related to the foreground maps mostly involve maps and spherical harmonic expansions, the statistical tests, implemented in this work,  ensure that the inpainted maps do not have any bias when compared with the authentic ones outside the masked area. \xedit{However, we plan to address in a future work what is the eventual bias introduced by  inpainting methodology by estimating the bi- and tri-spectrum of angular correlation.} 

 {Finally, both the  two DCNN methodologies implemented in this work are deterministic,  meaning that they  generate one single inpainting result for a given corrupted image. This makes very hard to perform a $\chi^2$ analysis on the inpainted images as  the pixel-pixel  covariance matrix cannot be defined with one sole inpainting case. 
However, a novel methodology based on DCNN has been proposed by \citet{Cai2019},the  Pluralistic Image Inpainting GAN (PiiGAN\footnote{https://github.com/vivitsai/PiiGAN}). \xedit{The generator in PiiGAN is designed such that it can generate multiple results from the contextual semantics of one image allowing us to perform  the $\chi^2$ analysis on inpainted maps.} We devote a future work to apply PiiGAN for the specific case of Galactic Foreground inpainting.} 
\section{Summary and Conclusions } 

\noindent In this work, we demonstrate three inpainting methodologies that can reproduce an underlying non-Gaussian signal without modifying the overall summary statistics of the signal itself. 

The first method (NN) has already been used in the literature and it is based on diffusing the pixels in the masked area with the average of nearest-neighbour pixels. We further adopte two novel techniques relying on DCNNs, namely DP and GAN. They are first introduced as a tool to inpaint natural images by the deep learning community.
We validate the three techniques on simulated data, and test them on   data-set with a wide range of SNRs. In addition, we show a real-life application by inpainting a map in regions where bright point-sources are detected. 
To evaluate the quality of inpainted results,  we adopted three summary statistics based on the pixel distribution, including the angular power spectra, and the first three Minkowski functionals.
We find that all techniques are able to reproduce the overall summary  statistics when applied to signal-only data. 

Inside the masked regions inpainted with NN, the results are smooth but lacks of finer details. Generally, because NN averages the neighboring pixels, a map inpainted with NN sharply transitions from the area outside the mask with sub-structures and noisy pixels to a very smooth one encoding only long and smooth modes inside the masked area. However, we did not notice any clear effect or bias due to NN inpainting on the statistical tests we adopted in this work. 

On the other hand, DP reconstructions on  images extracted from  signal and noise maps present a different  Minkowski functionals with respect to the ground-truth ones.  As pointed out at the end of Sect.\ref{subsec:fidelity}, this failure case of DP needs to be further investigated by means of a better tuning of hyper-parameters. 

 GAN has been demonstrated to be promising as it is able to statistically reproduce signal and noise maps, and it generates images visually very similar to the ground-truth on signal dominated maps. However, we have identified cases where it fails to produce high fidelity images in noise dominated maps. We plan to further investigated this in a future work by training GAN with more realistic dataset including several levels of SNR.

To our knowledge, this is the first time that GAN have been used to successfully generate high resolution intensity and polarization maps of  Galactic foreground polarization maps.
This approach opens up many possibilities of generating foreground maps using adversarial networks, which overcome the limitations of existing templates. 

In conclusion, we focused this work  on  Galactic foreground  emission motivated by the challenges in inpainting non-Gaussian signal and in dealing with Galactic (and Extra-galactic) foregrounds in CMB B-mode polarization studies. For the future work, we plan to apply similar techniques to different non-Gaussian  signals spanning from galaxy weak lensing to HI data, which are highly affected by foreground emission as well. 

\vspace{1cm}

\noindent \emph{Acknowledgements.} The authors are very thankful to Yuuki Omori and Ben Thorne for  carefully reading the manuscript.   They also  thank Alexandre Refregier, Nicoletta Krachmalnicoff, Ioannis Liodakis and Warren Morningstar for the fruitful discussions and useful comments shared during the development of this work. Some of the results in this paper have been derived using the HEALPIX \citep{2005ApJ...622..759G} package.


\bibliography{inpaint,methods}{}
\bibliographystyle{aasjournal}
\newpage 

%\appendix
\restartappendixnumbering
%\input{appendix}


\end{document}